\documentclass[nameyear,citesort,seceqn,dvips]{arxbj}
\usepackage{graphicx}

\aid{0}
\volume{17}
\issue{3}
\pubyear{2011}
\firstpage{845}
\lastpage{879}
\doi{10.3150/10-BEJ309}

\makeatletter

\newtheorem{prop}{Proposition}
\newtheorem{coro}{Corollary}
\newproclaim{defn}{Definition}
\newtheorem{lemma}{Lemma}

\newremark{exa}{Example}
\def\eqref#1{(\ref{#1})}

\newcommand{\nn}[0]{\hspace*{.7em}}
\newcommand{\n}[0]{\hspace*{.35em}}
\newcommand{\ful}{\frac{\nn\nn}{ \nn\nn}}
\newcommand{\fla}{\hspace*{-2pt} \prec\!\!\!\!\!\frac{\nn\nn}{\nn}}
\newcommand{\fra}{ \frac{\nn
\nn}{\nn}\!\!\!\!\! \succ\! }
\newcommand{\dal}{\hspace*{-1pt}\frac{\n}{\n}\frac{\hspace*{4pt}}{\hspace*{4pt}} \frac{\n}{\n}}

\makeatother

\begin{document}
\begin{frontmatter}

\title{Probability distributions with summary graph structure}
\runtitle{Summary graphs}

\begin{aug}
\author{\fnms{Nanny} \snm{Wermuth}\corref{}\ead[label=e1]{wermuth@chalmers.se}
\ead[label=u1,url]{http://www.math.chalmers.se/\textasciitilde wermuth}}
\runauthor{N. Wermuth}
\address{Mathematical Statistics at Chalmers,
University of Gothenburg, Chalmers Tv\"{a}rgata 3,
41296~G\"{o}\-teborg, Sweden. \printead{e1}; \printead{u1}}
\end{aug}

\received{\smonth{1} \syear{2009}}
\revised{\smonth{2} \syear{2010}}

%
\begin{abstract}
A set of independence statements may define the independence structure
of interest in a family of joint probability distributions. This
structure is often captured by a graph that consists of nodes
representing the random variables and of edges that couple node pairs.
One important class contains regression graphs. Regression graphs are a
type of so-called chain graph and describe stepwise processes, in which
at each step
single or joint responses are generated given the relevant explanatory
variables in their past. For joint densities that result after possible
marginalising or conditioning, we introduce summary graphs. These
graphs reflect the independence
structure implied by the generating process for the reduced set of
variables and they preserve the implied independences after additional
marginalising and conditioning. They can identify generating dependences
that remain unchanged and alert to possibly severe distortions due to
direct and indirect confounding.
Operators for matrix representations of graphs are used to derive these
properties of summary graphs and to translate them into special types
of paths in graphs.

\end{abstract}

%

\end{frontmatter}

\section{Motivation, some previous and some of the new results}

\subsection{Motivation}
Graphical Markov models are probability distributions defined for a
$d_V \times1$
random vector variable $Y_V$
whose component variables may be discrete, continuous or of both types and whose joint density
$f_V$ satisfies the independence statements specified directly by an associated graph as well as those
implied by the graph.
The set of all such statements is the independence
structure captured by the graph.

One such type of graph was introduced for sequences of regression by
Cox  and Wermuth (\citeyear{CoxWer93,CoxWer96}) for which special results have been derived
by Drton (\citeyear{Drton09}), Kang  and Tian (\citeyear{KangTian09}), Marchetti and Lupparelli (\citeyear{MarLup10}),
Wermuth and Cox (\citeyear{WerCox04}), Wermuth, Wiedenbeck and Cox (\citeyear{WerWieCox06}), Wermuth, Marchetti and Cox (\citeyear{WerMarCox09}),
Wermuth and Sadeghi (\citeyear{WerSad10}).

A regression graph consists of nodes, say in set $V$, that represent
random variables, and of edges that couple node pairs such that a
recursive order of the joint responses is reflected in the graph.
Associated discrete distributions have some desirable properties
derived by Drton (\citeyear{Drton09}).
Each defining independence constraint respects the given recursive
ordering of the joint responses; see Marchetti and Lupparelli (\citeyear{MarLup10}). This feature distinguishes regression graphs from all other
currently known types of chain graphs and permits one to model data from both interventional
and observational studies.

Because of this property, regression graphs are particularly well
suited to the study of effects of hypothesized causes on sequences of
joint responses; see Cox and Wermuth (\citeyear{CoxWer04}). More generally, they can model developmental
processes, such as in panel studies. These provide data on a group of
individuals, termed the `panel', collected repeatedly, say over years
or decades.
Often one wants to compare corresponding analyses with results in other studies
that have core sets of variables in common, but that have omitted some
of the variables or that were carried out for subpopulations.

It is an outstanding feature of regression graph models that their
implications can be derived after
marginalising over some variables, say in set $M$, or after
conditioning on others, say in set~$C$. In particular, graphs can be
obtained for node set $N=V\setminus\{C,M\}$ that capture precisely the
independence structure implied by a generating graph in node set $V$
for the distribution of
$Y_N$ given $Y_C$.

Such graphs are called independence-preserving, when they can be used
to derive the independence structure that would have resulted from the
generating graph
by conditioning on a larger node set $\{C,c\}$ or by marginalising over
a larger node set $\{M,m\}$.
Two types of such classes are known. One is the subclass of the much
larger class of MC graphs of \citet{Kost02}, which can be generated by
a regression graph in a larger node set. Another class contains the MAG's (maximal
ancestral graphs) of \citet{RicSpi02}. We speak of two corresponding
graphs if they result from a given generating graph relative to the
same conditioning and marginalising sets.

A third
class of this type is the summary graph of \citet{WerCoxPea94}. This
class is presented in the current paper in simplified form together
with proofs based on operators for binary matrix representations of the
graphs. In contrast to a MAG, a corresponding summary graph can be used to
identify those dependences of a given generating process for $Y_V$
with $V>N$
that remain undistorted in the corresponding MAG model for $Y_N$ given
$Y_C$ and those that may be severely distorted.
This is especially helpful at the planning stage of studies when
alternative sets $M$ and $C$ are considered given a hypothesized
generating graph in $V>N$. Annotated, undirected graphs of \citet
{Paz07}, for $C$ empty, serve a similar purpose.

The warning signals for distortions provided by summary graphs are
essential for understanding consequences of a given data generating
process with respect to
dependences in addition to independences. For this, some special
properties of the types of generating graph will be introduced as well
as specific requirements on the types of generating process. These lead
to families of distributions that are said to be generated over parent graphs.

\subsection{Some notation and concepts}

Some definitions for graphs are almost self-explanatory. If pair $i
\neq k$ of $V$ is coupled by a~directed edge such that an arrow starts
at node $k$ and points to node $i$, then $k$ is named a parent of $i$
and $i$ the offspring of $k$. For two disjoint subsets $\alpha$ and
$\beta$ of $V$, an $ik$-arrow, $i\fla k$, is said to point
from $\beta$ to $\alpha$ if the arrow starts at a node $k$ in $\beta$ and points to a~node $i$ in $\alpha$. Nodes other than
the endpoint nodes are the inner nodes of a path; only the inner nodes have to be distinct. For
three or more nodes, an $ik$-path connects the path endpoint nodes $i$ and $k$ by a sequence of
edges that couple its inner nodes. An $ik$-path with $i = k$ is a cycle.

An edge is regarded as a path without inner nodes.
Both a graph and a path are called directed if all its edges are
arrows. If all arrows of a directed $ik$-path point towards node~$i$,
then node $k$ is named an ancestor of $i$ and $i$ a descendant of $k$.
Such a path is also called a descendant--ancestor path.

Directed acyclic graphs form an important subclass of regression
graphs. They
arise from stepwise generating processes of exclusively univariate
response variables; see Section~\ref{sec2} below.
These graphs have no directed cycles.

As we shall show, two different types of undirected graph are
subclasses of regression graphs, named covariance graphs and
concentration graphs.
For joint Gaussian distributions, they give models for
zero constraints on covariances or on concentrations, respectively; see
\citet{WerCox98} and \eqref{overallg} and \eqref{indMRC} below.
To distinguish between them in figures, edges in concentration graphs
are shown as full lines,
$ i \ful k$, and in covariance graphs by dashed lines, $i \dal k$.

Separation criteria provide what is called the global Markov property
of a graph since it gives all independence statements that belong to
the graph's independence structure.

\begin{defn}\label{d1} A graph, consisting of a node set and of one or more edge
sets, is an independence graph if node pairs are coupled by at most one
edge and each missing edge corresponds to at least one independence
statement.
\end{defn}

Regression graphs and MAGs are independence graphs but, in general,
summary graphs, ancestral graphs and MC graphs are not, even with at
most one edge for each node pair;
see the discussion of Figure~\ref{fig3}(b) below.

The same graph theoretic notion of separation applies to both types of
undirected graph.
Let $\alpha$ and $\beta$ be two non-empty, disjoint subsets of their
node set $V$ and let $\{\alpha, \beta, m,c\}$ partition $V$, then we
write $Y_\alpha$ is conditionally independent of $Y_\beta$ given
$Y_c$ compactly as $\alpha\perp\!\!\!\hspace*{-1pt}\perp\beta|c$.
In a concentration graph, $\alpha$ is separated by $c$ from $\beta$
if every path from $\alpha$ to $\beta$ has a node in $c$, while in
the covariance graph, $\alpha$ is separated by $m$ from $\beta$ if
every path from $\alpha$ to $\beta$ has a node in $m$. Given
separation of $\alpha$ and $\beta$ by set $c$, a~concentration graph
implies $\alpha\perp\!\!\!\hspace*{-1pt}\perp\beta|c$; see \citet{Lau96}.
Given separation of $\alpha$ and~$\beta$ by set $m$, a covariance
graph implies $\alpha\perp\!\!\!\hspace*{-1pt}\perp\beta|c$; see \citet{Kau96},
who expresses
the result in a different but equivalent way.

When a graph is directed or contains different types of edge then its
separation criterion is more complex than the one for undirected graphs.
For directed acyclic graphs, there are several different separation
criteria that permit us to obtain all independence statements implied by
the graph; see \citet{MarWer09} for proofs of equivalence.

The criterion due to \citet{GeiVerPea90}, has been extended in almost
unchanged form by Koster (\citeyear{Kost02}) to the much larger class of MC
graphs. A path-based proof, due to \citet{Sadeghi09}, is for the
subclass of MC graphs that is of interest here, the MC graphs that can
be derived from a larger directed acyclic
graph. For summary graphs, see Lemma~\ref{l1} below.

A list of independence statements associated with the missing edges of
an independence graph gives a graph's pairwise Markov property.
Whenever it defines the graph's independence structure, then the
pairwise Markov property is said to be
equivalent to the global Markov property.

For all disjoint subsets $a,b,c,d$ of node set $V$, the following
general definitions are relevant, respectively, for combining pairwise
independences in covariance graph and in concentration graph models.

\begin{defn}\label{d2} The composition property is
\[
a\perp\!\!\!\hspace*{-1pt}\perp b |d \mbox{ and } a \perp\!\!\!\hspace*{-1pt}\perp c|d \mbox{
imply } a\perp\!\!\!\hspace*{-1pt}\perp bc|d  .
\]
\end{defn}

\begin{defn}\label{d3} The intersection property is
\[
a\perp\!\!\!\hspace*{-1pt}\perp b |cd \mbox{ and } a \perp\!\!\!\hspace*{-1pt}\perp c|bd \mbox{
imply } a\perp\!\!\!\hspace*{-1pt}\perp bc|d  .
\]
\end{defn}

Given these properties, the independence structure of interest in a
covariance or concentration graph model can be specified in terms
of independence constraints on a set of variable pairs. For general
searching discussions, see \citet{Daw79}, \citet{Pea88}, \citet{Lau96} and \citet{Stu05}.

Necessary and sufficient conditions under which discrete and Gaussian
distributions satisfy the intersection property  have been derived by
San Martin, Mochart  and Rolin (\citeyear{SanMar05}). They show in particular that of the commonly
specified sufficient conditions, some may be much too strong -- for
instance, requiring exclusively positive probabilities for discrete
distributions.
For joint Gaussian distributions, a positive definite joint covariance
matrix is sufficient. In both cases, no component of the involved
random variables is degenerate.

\begin{defn}\label{d4}
A family of joint distributions is said to vary fully if its random
variables contain no degenerate components and it satisfies
the intersection property.
\end{defn}

\begin{defn}\label{d5}
In families of joint distributions with the composition property,
pairwise independent variables are also mutually independent.
\end{defn}

For families of joint distributions with the composition property, in
which a regression graph with a complete concentration graph captures
the independences of interest, the global and the pairwise Markov
property are equivalent; see also \citet{KangTian09}.

For a long time, only the family of Gaussian distributions was known to
satisfy both the composition and the intersection property
provided it varies fully. Under the same type of constraint, this is
now known to hold for the special family of distributions in symmetric
binary variables introduced by Wermuth, Marchetti and Cox (\citeyear{WerMarCox09}). More important, as
we shall see, it holds for families generated over so-called parent
graphs.

The notion of completeness has been introduced and studied in quite
different contexts [see \citet{LehSch55}; \citet{Brown86}, Theorem
2.12; and Mandelbaum and R\"{u}schendorf \citeyear{ManRue87}]. It means that the joint family of
distribution of vector variable $Y$ is such that
a zero expectation of any function $g(y)$ implies that the function
itself is zero with probability one, that is, almost surely (a.s.).

\begin{defn}\label{d6} Let $f(y)$ denote the density of a member of a complete
family of distributions
and $g(y)$ be some function of $Y$. Then it holds that
\[
\int g(y) f(y)   \,\mathrm{d} y= 0\quad \Longrightarrow\quad g(y) =0\qquad \mbox{a.s.}
\]
\end{defn}

For any trivariate family of distributions with precisely two
associated variable pairs, say $(Y_1, Y_2)$ and $(Y_1, Y_3)$, but
$2\perp\!\!\!\hspace*{-1pt}\perp
3$, completeness of the joint distribution is sufficient to conclude
that $Y_2$ is conditionally dependent on $Y_3$ given $Y_1$. This
follows from Corollary~3 of \citet{WerCox04} and properties of
completeness. In this situation, the generating graph
\[
\vspace*{-2pt}
2 \fra1\fla3
\vspace*{-2pt}
\]
is inducing a $2, 3$-edge in the summary graph obtained by conditioning on
node 1 and a~non-vanishing conditional association for $Y_2 , Y_3$
given $Y_1$.

In Section \ref{sec2}, we define parent graphs as directed acyclic graphs with
special properties and corresponding types of stepwise generating processes
such that edge-inducing paths are also association-inducing. The
families of distributions generated over parent graphs
and the members of the families satisfy the intersection and the
composition property in addition to the general laws of probability
that govern independences in any joint family of distribution; for a
discussion of the latter see \citet{Stu05}.

\subsection{Definition and construction of summary
graphs}\label{sec1.3}

In contrast to MC graphs and MAGs, regression graphs are not closed
under marginalising and conditioning, that is, one can get from a given
regression graph outside the class of regression graphs after
marginalising and conditioning as illustrated with Figure~\ref{fig3} below. But
the graph resulting in this way
from any regression graph is always within the class of summary graphs.
This explains partly why we study the larger class of summary
graphs.\looseness=-1

\begin{defn}\label{d7} A summary graph, $G^N_{\mathrm{sum}}$, has node set
$N$, which
consists of disjoint subsets $u,v$, ordered as $(u,v)$. Within $u$, the
graph has a mixture of a directed acyclic graph and of a covariance
graph and, within $v$, it has a concentration graph. Between $u$ and
$v$, only arrows point from $v$ to $u$.
\end{defn}

The notions of parents, offsprings, ancestors and descendants remain
unchanged in a~summary graph compared to a directed acyclic graph.
As will be shown, every summary graph in node set $N$ can be generated
from a directed acyclic graph in node set $V=\{\mbox{\Large{$\mbox{$\circ$}$}}\}$ by
conditioning on $C=\{\raisebox{-2pt}{\mbox{\includegraphics{309i01}}}\}$ and marginalising over
 $M=\{\raisebox{-3pt}{\mbox{\includegraphics{309i02}}}\}$
so that $N=V\setminus\{C,M\}$. This graph is denoted by $G^{V\setminus
  [C,M]}_{\mathrm{sum}}$, an associated density by $f_{N|C}$ that
results from $f_V$, the given density of the generating graph, which factorizes
according to this graph; see~\eqref{fact} below.

The density $f_{N|C}$ may concern discrete, continuous or mixed
variables, as implied by~$f_V$. It has a factorization according to
$(u,v)$ that is written compactly in terms of
node sets as
%
\begin{equation}\label{eqfacts}
 f_{N|C}=f_{u|v C}f_{v |C}.
\end{equation}
In the larger generating graph in node set $V$, every node in $v$ and
no node in $u$ is an ancestor of the conditioning set $C$. Thus, each
component of $Y_v$ has been generated~befo\-re~$Y_u$; see Figure~\ref{fig2} for an example.

\begin{figure}[t]

\includegraphics{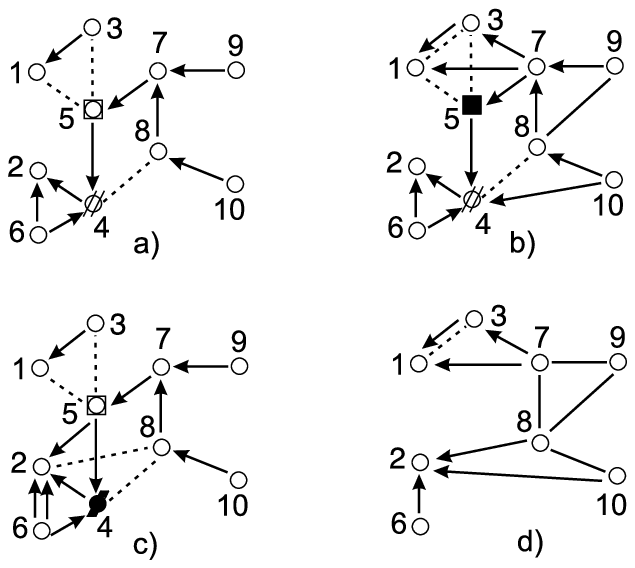}
\vspace*{-3pt}
\caption{\textup{(a)} A summary graph with node 4 to be marginalised over and
node 5 to be conditioned on, \textup{(b)}~the graph of \textup{(a)} including edges induced
for conditioning on node 5, \textup{(c)} the graph of \textup{(a)} including edges induced
for marginalising over node 4, \textup{(d)} $G^{N\setminus [5,4]}_{\mathrm{sum}}$.}
\label{fig1}
\vspace*{-5pt}
\end{figure}

\begin{figure}

\includegraphics{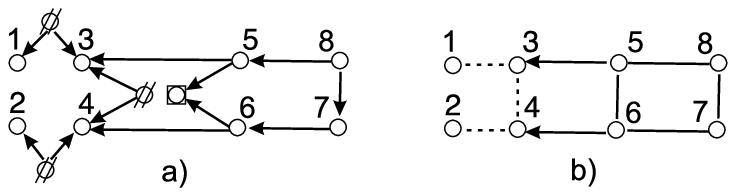}
\vspace*{-3pt}
\caption{\textup{(a)} A directed acyclic graph generating \textup{(b)} a summary graph
without semi-directed cycles; $u=\{1,2,3,4\}$ and $v=\{5,6,7,8\}$.}
\label{fig2}
\vspace*{-5pt}
\end{figure}

Figures~\ref{fig1}--\ref{fig3} illustrate how summary graphs may be generated. For this,
the stepwise construction of a summary graph by marginalising over $m=\{
t\}$ or conditioning on $c=\{s\}$ in $G^N_{\mathrm{sum}}$ is given
in Table~\ref{fig1}.

\begin{table}[b]
\tablewidth=305pt
\caption{Types of induced edge when each of $m$ or $c$
contains a single node in $G^N_{\mathrm{sum}}$}\label{tab1}
\begin{tabular*}{305pt}{@{\extracolsep{4in minus 4in}}lcccc@{}}
\hline
\multicolumn{5}{c@{}}{Types of induced edge when
marginalising over the common neighbor node $t$}
\\
\hline
& \hspace*{1pt}$t   \fra  \mbox{\Large{$\mbox{$\circ$}$}}$& \hspace*{0,3pt}$  t  \ful  \mbox{\Large{$\mbox{$\circ$}$}}  $& \hspace*{0,5pt}$t   \fla
\mbox{\Large{$\mbox{$\circ$}$}}$& \hspace*{0,5pt}$t   \dal  \mbox{\Large{$\mbox{$\circ$}$}}$
\\[-6pt]
&\multicolumn{4}{c@{}}{\hrulefill}\\
$\mbox{\Large{$\mbox{$\circ$}$}}  \fla  t$ & $\mbox{\Large{$\mbox{$\circ$}$}}  \dal  \mbox{\Large{$\mbox{$\circ$}$}}$ & $\mbox{\Large{$\mbox{$\circ$}$}}
\fla  \mbox{\Large{$\mbox{$\circ$}$}}$ & $\mbox{\Large{$\mbox{$\circ$}$}}  \fla  \mbox{\Large{$\mbox{$\circ$}$}}$& $\mbox{\Large{$\mbox{$\circ$}$}}  \dal
\mbox{\Large{$\mbox{$\circ$}$}}$
\\
\hspace*{1pt}$\mbox{\Large{$\mbox{$\circ$}$}}\ful  t $& $\cdot$ & \hspace*{-0,1pt}$\mbox{\Large{$\mbox{$\circ$}$}}  \ful  \mbox{\Large{$\mbox{$\circ$}$}}$& $\mbox{\Large{$\mbox{$\circ$}$}}
\ful  \mbox{\Large{$\mbox{$\circ$}$}}$& $ \mbox{\Large{$\mbox{$\circ$}$}}  \fra  \mbox{\Large{$\mbox{$\circ$}$}}$\\
\hline
\end{tabular*}
\begin{tabular*}{305pt}{@{\extracolsep{4in minus 4in}}lcccc@{}}
\multicolumn{5}{@{}l}{and types of induced edge when conditioning on the common neighbor node
$s$ or on}\\
\multicolumn{5}{@{}l}{one of the descendants of $s$}\\[-6pt]
&\multicolumn{3}{c@{}}{\hspace*{-6pt}\rule{196pt}{0.5pt}}&\\
&& \hspace*{0,7pt}$s   \fla  \mbox{\Large{$\mbox{$\circ$}$}}$& \hspace*{0,6pt}$s   \dal  \mbox{\Large{$\mbox{$\circ$}$}}$
&\\[-6pt]
&\multicolumn{3}{c@{}}{\hspace*{-6pt}\rule{196pt}{0.5pt}}&\\
&$\mbox{\Large{$\mbox{$\circ$}$}}  \fra  s$ & $\mbox{\Large{$\mbox{$\circ$}$}}  \ful  \mbox{\Large{$\mbox{$\circ$}$}}$
&$ \mbox{\Large{$\mbox{$\circ$}$}}  \fra\mbox{\Large{$\mbox{$\circ$}$}}$&
  \\
&$\mbox{\Large{$\mbox{$\circ$}$}}  \dal  s $& $\cdot$ & \hspace*{-0,1pt}$\mbox{\Large{$\mbox{$\circ$}$}}  \dal  \mbox{\Large{$\mbox{$\circ$}$}}$& \\[-6pt]
&\multicolumn{3}{c@{}}{\hspace*{-6pt}\rule{196pt}{0.5pt}}&\\
\multicolumn{5}{@{}l}{where the $\cdot$ notation indicates a symmetric
entry.}\\\hline
\end{tabular*}
\end{table}

\begin{figure}[t]

\includegraphics{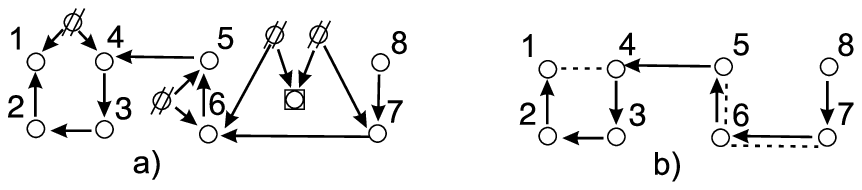}
\vspace*{-3pt}
\caption{\textup{(a)} A directed acyclic graph generating \textup{(b)}~a summary graph
with $v$ as the empty set and several semi-directed cycles;
the $4, 4$-path with inner nodes $1,2,3$, the $6, 6$-path via inner node $5$
and the double edge for (6, 7).}
\label{fig3}
\vspace*{-4pt}
\end{figure}

If a node $t$ is coupled with both of the nodes $i$ and $k$, then $t$
is said to be their common neighbor.
In two-edge paths, the inner node is named a collision node for
\[
\mbox{\Large{$\mbox{$\circ$}$}}\fra\mbox{\Large{$\mbox{$\circ$}$}}\fla\mbox{\Large{$\mbox{$\circ$}$}},\qquad\mbox{\Large{$\mbox{$\circ$}$}}
\fra\mbox{\Large{$\mbox{$\circ$}$}}\dal
\mbox{\Large{$\mbox{$\circ$}$}},\qquad\mbox{\Large{$\mbox{$\circ$}$}}\dal\mbox{\Large{$\mbox{$\circ$}$}}\dal\mbox{\Large{$\mbox{$\circ$}$}},
\]
and a transmitting node, otherwise. A path for which all inner nodes
are collision nodes is a collision path and a path for which
all inner nodes are transmitting nodes is a~transmitting
path.

Table~\ref{tab1} is taken from \citet{WerCoxPea94}. In the \hyperref[app]{Appendix} here, we show
that the types of edge are self-consistent when they are induced using
Table~\ref{tab1}. The table implies in particular that a collision node is
edge-inducing by conditioning on it while a~transmitting node is
edge-inducing by marginalising over it.

Let a summary graph, $G^N_{\mathrm{sum}}$, be given and nodes
$s\neq t$ of $N$ be
selected. Suppose one intends to marginalise over node $t$ and to
condition on node $s$ and $d_s$ denotes the ancestors of $s$ within $u$
of $G^N_{\mathrm{sum}}$.
Then, a new summary graph in node set $N'=N\setminus\{s,t\}$ results
by using the procedure given in the following Proposition~\ref{p1}.
The graph $G^{N'}_{\mathrm{sum}}$ has its concentration graph in
$v'=v\setminus\{
s,t\}$ whenever both nodes are in $v$, in $v'=v\setminus\{ s\}$ for
only $s$ in $v$, in $v'=\{v\setminus\{t\}, d_s\}$ for only $t$ in $v$
and in $v'=\{v, d_s\}$ for both nodes in $u$.

\begin{prop}[(Generating a summary graph from
$\bolds{G}^{\bolds{N}}_{\mathrm{\mathbf{sum}}}$ by operating on at most two nodes)]\label{p1}  From
$G^N_{\mathrm{sum}}$, the
independence-preserving
summary graph $G^{N\setminus [s,t]}_{\mathrm{sum}}$
is generated, with $t$ the marginalising node and $s$ the conditioning
node, by
inducing edges as prescribed in Table~\ref{tab1}:
\begin{enumerate}[(1)]
\item[(1)] first for the neighbors of $t$, second for the neighbors of $s$ and
of all
of its ancestors, ignoring in the second step edges
involving $t$,

\item[(2)] changing each edge present within $v'$ into a full line and each
edge present between $u'$ and $v'$ into an arrow
pointing from
$v'$ to $u'$,

\item[(3)] keeping for each node pair of several edges that are of the same
kind just one and deleting all nodes and edges
involving $s$ or $t$.
\end{enumerate}
\end{prop}

Section~\ref{sec3} contains proofs in terms of operators for matrix
representations of graphs. The proofs imply for any node subset $\{m,
c\}$ of $N$ that $G_{\mathrm{sum}}^{N\setminus  [\varnothing,
m]}$ may be
derived before conditioning on set $c$, or $G_{\mathrm
{sum}}^{N\setminus [c,
\varnothing]}$ before marginalising over set $m$ and that within sets
$c$ or $m$ any order of the nodes can be chosen. In particular, in step
(1) one may first work on the neighbors of $s$ and of all of its
ancestors and second
on the neighbors of $t$, ignoring edges involving $s$ in this second step.

The matrix formulations
lead more directly to $G_{\mathrm{sum}}^{N\setminus [c, m]}$,
but Proposition~\ref{p1}
gives an algorithm for operating on one node at a time.\vadjust{\goodbreak}
It is also helpful for small graphs, as
illustrated with Figures~\ref{fig1}--\ref{fig3}. Proposition~\ref{p1} implies that no
coupled pair ever gets uncoupled and that
the two types of path that may occur when constructing a summary graph
are replaced in $G^{N\setminus [j,t]}_{\mathrm{sum}}$:
\begin{eqnarray*}
&\mbox{\Large{$\mbox{$\circ$}$}}\fra\mbox{\Large{$\mbox{$\circ$}$}}\ful\mbox{\Large{$\mbox{$\circ$}$}}
\quad\mbox{by}\quad
\mbox{\Large{$\mbox{$\circ$}$}}\ful\mbox{\Large{$\mbox{$\circ$}$}}\ful\mbox{\Large{$\mbox{$\circ$}$}},&
\\
&\mbox{\Large{$\mbox{$\circ$}$}}\dal\mbox{\Large{$\mbox{$\circ$}$}}\ful\mbox{\Large{$\mbox{$\circ$}$}}\quad\mbox{by} \quad\mbox{\Large{$\mbox{$\circ$}$}}\fla\mbox{\Large{$\mbox{$\circ$}$}}
\ful\mbox{\Large{$\mbox{$\circ$}$}}.&
\end{eqnarray*}

The starting summary graph of Figure~\ref{fig1} is in \ref{fig1}(a). For $j=5$ and
$t=4$, Figure~\ref{fig1}(b) shows the edges induced by operating first on $j$,
Figure~\ref{fig1}(c) those induced by operating first on $t$ and Figure~\ref{fig1}(d) displays $G^{N\setminus [5,4]}_{\mathrm{sum}}$.

By construction, a summary graph contains
no directed cycle, but possibly semi-directed cycles. These are
direction-preserving cycles containing at least one undirected edge;
see, for instance, nodes $1,2,3,4$ of Figure~\ref{fig3}(b).

\begin{coro}[(Regression graphs and summary graphs)]\label{c1}
A regression graph is a~summary graph without semi-directed cycles.
\end{coro}

In contrast to a summary graph, a regression graph is an independence
graph that has at most one edge coupling any node pair; compare Figures~\ref{fig2}(b) and \ref{fig3}(b).
Figure~\ref{fig2}(b) shows a regression graph generated from a directed
acyclic graph and Figure~\ref{fig3}(b) a~summary graph with semi-directed
cycles.

By replacing each dashed $ik$-edge by an $ik$-path $i\fla\raisebox{-3pt}{\mbox{\includegraphics{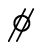}}}\fra k$, every summary graph has a virtual generating directed acyclic graph
for the nodes within $u$ even though a~dashed line might actually have
been generated by over-conditioning, that is, by including an offspring
in the conditioning set of two of its parents; see, for example,
$\raisebox{-3pt}{\mbox{\includegraphics{309i02.eps}}}\fra\raisebox{-1pt}{\mbox{\includegraphics{309i01}}}\fla\raisebox{-3pt}{\mbox{\includegraphics{309i02}}}$ as the inner nodes of the 6, 7 path
in Figure~\ref{fig3}(a).

Similarly, cycles in four or more nodes within $v$ may be generated
from a larger directed acyclic graph by including additional nodes,
$\raisebox{-1pt}{\mbox{\includegraphics{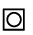}}}$, in appropriate ways; see \citet{CoxWer00}. The summary
graph in node set $N$ is uniquely defined if generated from a directed
acyclic graph in node set $V$ for given sets $M,C$, but typically many
different directed acyclic graphs, in node sets larger than $N$, may
lead to the same summary graph.

\subsection{Independence interpretation of summary graphs}

A criterion to decide whether a given summary graph, $G^{V\setminus
[C,M]}_{\mathrm{sum}}$, implies $\alpha\perp\!\!\!\hspace*{-1pt}\perp\beta|
c  C$ is given next.
For this, the node set $N$ is partitioned as $N=\{\alpha,\beta,c, m\}
$, where only subsets~$c$ or $m$
may be empty.
\begin{lemma}[(Path criterion
for the global Markov property {[\citet{Kost02}, \citet{Sadeghi09}]})]\label{l1}  The graph
$G^{V\setminus[C,M]}_{\mathrm{sum}}$ implies $\alpha\perp\!\!\!
\perp\beta|c  C$ if and
only if it has no $ik$-path between $\alpha$ and $\beta$ such that
every inner collision node
is in $c$ or has a descendant in $c$ and every other inner node is
outside~$c$.
\end{lemma}

In addition to the directly described path, Lemma~\ref{l1} specifies many
special types of forbidden path. We name a path of $n>2$ nodes an
$a$-line path if all inner nodes are within set $a$.\vadjust{\goodbreak}
The marginalising set for $\alpha\perp\!\!\!\hspace*{-1pt}\perp\beta|c  C$ in
$G^{V\setminus
[C,M]}_{\mathrm{sum}}$ is implicitly defined by $m=N\setminus\{
\alpha, \beta, c
\}$. Then, in $G^{V\setminus[C,M]}_{\mathrm{sum}}$,
there should be for node $i$ in $\alpha$ and node $k$ in $\beta$ no
$ik$-edge, no $m$-line transmitting $ik$-path, no $c$-line collision
$ik$-path and
no $ik$-path with all inner transmitting nodes in $m$ and all inner
collision nodes in $c$.

\begin{coro}[(Active $\bolds{ik}$-paths)] \label{c2} An $ik$-path in
$G^N_{\mathrm{sum}}$ is
active relative to $[c, m]$ if and only if it is an $ik$-edge or every
inner transmitting node is in $m$ and every inner collision node is in
$c$ or has a descendant in $c$.
\end{coro}

If an active $ik$-path relative to $[c, m]$ has uncoupled endpoints,
the path is closed by an $ik$-edge in $G^ {N\setminus[c,
m]}_{\mathrm{sum}}$. If
an active $ik$-path has coupled endpoints, the path is edge-inducing in
the construction process of $G^ {N\setminus [c, m]}_{\mathrm
{sum}}$. Thus, we
sometimes replace `active' by the more concrete term `edge-inducing'.

Figure~\ref{fig2}(b) represents a regression graph, hence each missing edge
corresponds to at least one independence statement. This contrasts with
Figure~\ref{fig3}(b), which has semi-directed cycles
and no independence statement is implied
for pairs $(1,5)$, $(5,7)$, $(5,8)$, $(6,8)$. For pair $(1,5)$, we give more
detailed arguments.

In the graph of Figure~\ref{fig2}(b), node 3 has no descendants and is an
inner collision node in every path connecting 1 and 5. Hence, when
node 3 is marginalised over, $1 \perp\!\!\!\hspace*{-1pt}\perp5|C$ is implied.
In the graph of
Figure~\ref{fig3}(b), pair $(1,5),$ is connected by a descendant--ancestor
path with inner nodes in $\{2,3,4\}$. Therefore, a 1, 5-edge is induced
by marginalising over nodes $2,3,4$ and hence $1\perp\!\!\!\hspace*{-1pt}\perp5|C$ is
not implied.
A $1, 5$-edge is induced by conditioning on node 4 or on any of its
descendants in $\{1,2,3\}$ so that $1\perp\!\!\!\hspace*{-1pt}\perp5|c  C$
is not implied, $c \neq\varnothing$.

\begin{figure}[t]

\includegraphics{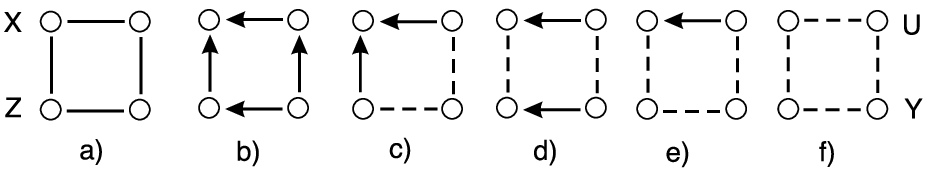}

\caption{Important special cases of summary graphs. The two pairs
$X,Y$ and $Z,U$ are constrained given $Y_C$; with $X\perp\!\!\!\hspace*{-1pt}\perp
Y|ZU$ in
\textup{(a)--(c)}, with $X\perp\!\!\!\hspace*{-1pt}\perp Y|U$ in \textup{(d)}, \textup{(e)} and with $X\perp\!\!\!\hspace*{-1pt}\perp Y$ in~\textup{(f)}; with $Z\perp\!\!\!\hspace*{-1pt}\perp
U$ in \textup{(c)}, \textup{(e)}, \textup{(f)}, with $Z\perp\!\!\!\hspace*{-1pt}\perp U|Y $ in \textup{(b)}, \textup{(d)} and with
$Z\perp\!\!\!\hspace*{-1pt}\perp U|XY$ in \textup{(a)}.}
\label{fig4}
\end{figure}

Figure~\ref{fig4} shows special cases of summary graphs, noting that $C$ and one
of $u$, $v$ may be empty sets.
Figure~\ref{fig4} shows that summary graphs cover
all six possible combinations of independence constraints on two
non-overlapping pairs of four variables $X,Z,U,Y$.
Substantive research examples with well-fitting data to linear models
of Figure~\ref{fig4} have been given by Cox and Wermuth (\citeyear{CoxWer93}) to the concentration
graph in Figure~\ref{fig4}(a), the directed acyclic graph in Figure~\ref{fig4}(b),
the graph of seemingly unrelated regression graph in Figure~\ref{fig4}(d) and the
covariance graph in Figure~\ref{fig4}\textup{(f)}.

\subsection{Markov equivalence}
The notion of Markov equivalence is important, because for any given
set of data one cannot
distinguish between two Markov equivalent graph models on the basis of
goodness-of-fit tests.

\begin{defn}\label{d8} Two different graphs in node set $N$ are Markov equivalent
if they capture the same independence structure.
\end{defn}

Since a different set of two independence statements is associated with
each of the graphs in Figure~\ref{fig4}, none of the six graphs are Markov equivalent.

Known conditions, under which a concentration graph or a covariance
graph is Markov equivalent to a directed acyclic graph, may be proven
by orienting the graphs, that is, by changing each edge present into an
arrow. The same type of argument can be extended to other independence
graphs such as to regression graphs; see also Proposition~\ref{p2} below. For
this, we need a few more definitions for graphs.

For $a \subset N$, the subgraph induced by $a$ is obtained by keeping
all nodes in $a$ and all edges coupling nodes within $a$.
A subgraph induced by three nodes that has two edges is named a \textsf
{V}-configura\-tion or simply a {\sf V}. A path is said to be chordless
if each inner node forms a \textsf{V} with its two neighbors.

For \textsf{V}'s of a regression graph that are collision paths with
endpoints $i$ and $k$, the inner node is excluded
from the conditioning set of every independence statement for $Y_i,
Y_k$ implied by the graph. In contrast, for \textsf{V}'s of a
regression graph that are transmitting paths, the inner
node is included in the conditioning set of every independence
statement for $Y_i, Y_k$ implied by the graph. Thus, the independence structure
of the graph would be changed whenever any collision \textsf{V} were
exchanged by a transmitting \textsf{V}.

A concentration graph with a chordless 4-cycle, as in Figure~\ref{fig4}(a),
or with any larger chordless cycle, is not Markov equivalent to a
directed acyclic graph; see \citet{Dir61}
and \citet{Lau96}.
The reason is that it is impossible to orient the graph, that is, to
replace each edge by an arrow, without obtaining either a directed
cycle or at least one collision
\textsf{V}.

Similarly, a covariance graph is not Markov equivalent to a directed
acyclic graph if it contains a chordless collision path in four nodes;
see \citet{PeaWer94}. The reason is that it is impossible to orient
each edge without obtaining at least one transmitting
\textsf{V}. There are the following three types of chordless collision
paths in four nodes in a regression graph:
\[
\mbox{\Large{$\mbox{$\circ$}$}}\fra\mbox{\Large{$\mbox{$\circ$}$}}\dal\mbox{\Large{$\mbox{$\circ$}$}}\fla\mbox{\Large{$\mbox{$\circ$}$}},
\qquad\mbox{\Large{$\mbox{$\circ$}$}}
\dal
\mbox{\Large{$\mbox{$\circ$}$}}\dal\mbox{\Large{$\mbox{$\circ$}$}}\fla\mbox{\Large{$\mbox{$\circ$}$}},
\qquad\mbox{\Large{$\mbox{$\circ$}$}}\dal\mbox{\Large{$\mbox{$\circ$}$}}
\dal
\mbox{\Large{$\mbox{$\circ$}$}}\dal\mbox{\Large{$\mbox{$\circ$}$}}.
\]
The next result in Proposition~\ref{p2} explains why, in general,
three types of edge are needed after marginalising and conditioning in
a directed acyclic graph.

\begin{prop}[(Lack of Markov equivalence)]\label{p2}
If a regression graph contains a~chordless collision path in four
distinct nodes or a chordless cycle in $n \geq4$ nodes within $v$,
then it is not Markov equivalent to any directed acyclic graph in the
same node set.
\end{prop}

\begin{pf} It is impossible to orient the graph with any one of the
above chordless collision paths in four nodes into edges of a directed
acyclic graph without switching between the two types of inner nodes in
at least one \textsf{V}, that is, between a collision and a
transmitting node. And, for the chordless cycle in $n \geq4$ nodes,
the above result for concentration graphs due to Dirac applies.
\end{pf}

Currently, one knows how to generate three types of
independence-preser\-ving graphs from a given directed acyclic graph in
node set $V$ for the same disjoint subsets $M$ and $C$ of $V$. In an MC
graph, four types of edge may occur in combination,
$i \fla k$, $i\fra k$, $i\dal k$ and $i\ful k$. A~summary graph may
have only one type of double edge, $i\matrix{\fla\vspace*{-8pt}\cr \dal\vspace*{1pt}} k$ and three types of
single edges, $i\fla k$, $i\dal k$ and $i\ful k$, while the
maximal ancestral graph is an independence graph with up to three types
of single edges, $i\fla k$, $i\dal k$ and $i\ful k$, where,
traditionally, the edge $i\dal k$ is drawn as a double-headed arrow.
For proofs of Markov equivalence of the three corresponding types of
graphs, see \citet{Sadeghi09}. In Section~\ref{sec3.6} below,
the unique MAG corresponding to a given summary graph is constructed.

\subsection{Families of distribution generated over parent graphs}

A distribution and its joint density $f_V$ is said to be generated over
a directed acyclic graph whenever $f_V$ factorizes recursively into
univariate conditional densities that satisfy the independence
constraints specified with the graph. Any full ordering of $V$ is
compatible with a given directed acyclic graph if, for each node $i$,
all ancestors of $i$ are in $\{i+1, \ldots, d_V\}$. The set of parent nodes
of $i$ is denoted by $\operatorname{par}_i$.

For $V=(1,\dots, d_V)$ specifying a compatible ordering
of node set $V$, a defining list of constraints for a directed acyclic
graph is
%
\begin{equation}\label{basic-ind}
f_{i|i+1, \dots, d_V}=f_{i|\operatorname{par}_i}\quad \iff\quad
i \perp\!\!\!\hspace*{-1pt}\perp\{
i+1,\dots, d_V\}\setminus\operatorname{par}_i| \operatorname
{par}_i
\end{equation}
and the factorization of the density generated over the graph is
%
\begin{equation}\label{fact}
f_V = \prod_{i = 1}^{d_V} f_{i|\operatorname{par}_i}.
\end{equation}

To generate $f_V$ recursively, one can take any compatible ordering of $V$.

\begin{defn}\label{d9} For a recursive generating process of $f_V$, one
starts with the marginal density $f_{d_V}$ of $Y_{d_V}$, proceeds with
the conditional density of $Y_{d_{V}-1}$ given $Y_{d_V}$, continues to
$f_{i|i+1, \dots, d_V}$ and ends with the conditional density of
$Y_{1}$ given $Y_{2}, \ldots, Y_{d_V}$.
\end{defn}

To let a directed acyclic graph represent one of such recursive generating
processes, the graph is to capture both independences and dependences.

\begin{defn} A directed acyclic graph, with a given compatible ordering
of $V$, is edge-minimal for $f_V$ generated over it if
\[
f_{i|\operatorname{par}_i} \neq f_{i| \operatorname{par}_i\setminus
l}\qquad \mbox{for each } l\in\operatorname{par}
_i .
\]
\end{defn}

Under this condition of edge-minimality of the generating graph for $f_V$,
all relevant explanatory variables are included for each $Y_i$ and no
edge can be removed from the graph without changing the independence
statements satisfied by $Y_i$ given its past, $\operatorname{pst}_i=\{
i+1, \dots,
d_V\}$.

An edge-minimal graph may represent a research hypothesis in a given
substantive context. For such a hypothesis, those dependences are
considered that are strong enough to be of substantive interest while
others are translated into independence statements;
see \citet{WerLau90}.

\begin{defn}
A recursive generating process of $f_V$ in the order $V=(1, \dots,
d_V)$ is said to
consist of freely chosen components $Y_i$ if each $Y_i$ can be discrete
or continuous and the parameters of $f_{i|\operatorname{pst}_i}$ are variation independent of those of $f_{\operatorname{pst}_i}$.
The form of the family of distribution of~$Y_i$ given $Y_{\operatorname
{pst}_i}$ may
be of any type.
\end{defn}

For exponential families of distributions, variation-independent
factorizations of $f_{i, \operatorname{pst}_i}=f_{i|\operatorname
{pst}_i}f_{\operatorname{pst}_i}$ coincide
with the notion of a cut given by Barndorff-Nielsen (\citeyear{Barn78}), page 50. These types
of factorization imply that the overall likelihood function can be
maximized by maximizing each factor $f_{i|\operatorname{pst}_i}$ separately.

In families of distribution with $f_V$ consisting of freely chosen
components that satisfy the defining independences \eqref{basic-ind}
of the given graph,
some further constraints on each $f_{i|\operatorname{par}_i}$ are
possible such as
no-higher-order interactions or such as requiring $Y_i$ to have
dependences of equal strength on several of its explanatory variables,
that is, on several components of $Y_{\operatorname{par}_i}$.
Excluded are, for instance, constraints across conditional distributions, such as dependences
of~$Y_i$ on some of $Y_{\operatorname{par}_i}$ to be equal to those of
$Y_k$ on some
of $Y_{\operatorname{par}_k}$.


Freely chosen components $Y_i$ are in general incompatible with
distributions that are to be faithful to a generating directed acyclic
graph. The notion was introduced by Spirtes, Glymour and Scheines
(\citeyear{Spirtesetal93}). It means that the independence structure of $f_V$ coincides
with the independence structure captured by the graph and it leads in
general to complex constraints on the parameter space for distributions
generated over parent graphs; see
Figure~1 of  \citet{WerMarCox09} for a simple example with three binary
variables. In contrast, variation independence permits special
constellations of parameter values that may lead to independences in
$f_V$ that are additional to those implied by the graph.

For research hypotheses, defined in terms of recursive constraints on
the independence structure and on dependences of $f_V$, appropriate
specifications and resulting properties can now be given. For this,
only connected graphs are considered, those with each node pair
connected by at least one path.\vspace*{2pt}

\begin{defn}\label{d12} A connected, directed, acyclic graph is named a parent
graph, $G_{\mathrm{par}}^V$ when one ordering of its node set
$V=(1,\dots, d_V)$
is given for the recursive generating process of $f_V$ and it is
edge-minimal for $f_V$.\vspace*{2pt}
\end{defn}

\begin{defn} A family of distributions is said to be generated over a
given parent graph if it varies fully and
each component of $f_V$ is freely chosen in the recursive generating
process of $f_V$.\vspace*{2pt}
\end{defn}

\begin{prop}[(General properties of families of distribution
generated over $\bolds{G}_{\mathbf{par}}^{\bolds{V}}$)] A family of distributions
generated over
$G_{\mathrm{par}}^V$ and each of its members satisfies
the intersection and the composition property. Every $ik$-path present
in $G_{\mathrm{par}}^V$ that induces an $ik$-edge by marginalising or
conditioning is also association-inducing for $Y_i, Y_k$.\vspace*{2pt}
\end{prop}

\begin{pf} The intersection property holds by the definition of
fully varying distributions. The composition property
holds by the definition of a parent graph since pairwise independences
without mutual independence cannot
result for edge-minimal, connected graphs that are directed and
acyclic. More precisely, let $ i < k$, and $c, d$ be disjoint subsets
of $\operatorname{pst}_i\setminus k$, then both of
$i \perp\!\!\!\hspace*{-1pt}\perp c |d$ and $k\perp\!\!\!\hspace*{-1pt}\perp c|d$ can be in the defining
list of independences only if the statement $i \perp\!\!\!\hspace*{-1pt}\perp c|kd$
is also
satisfied. In this
case, $f_{ikc|d} = f_{i|kd}f_{k |d} = f_{ik|d}$ so that $ik \perp\!\!\!
\perp c| d$
is implied. Finally, edge-minimality of a connected $G_{\mathrm
{par}}^V$ and
freely chosen densities $f_{i|\operatorname{pst}_i}$ assure that each
edge-inducing
path is also association-inducing.
\end{pf}

Excluded are incomplete families of distributions in which the
independence statement associated with each {\sf V} is not unique. For
instance, for an uncoupled node pair $i,k$ with transition \textsf{V},
$i\fla j \fla k$ and $\gamma\subseteq\operatorname{pst}_k$, it is
impossible that
\[
\int f_{ij|\gamma} f_{jk|\gamma}/f_{j|\gamma}  \,\mathrm{d} y_j = f_{i|\gamma
} f_{k|\gamma},\quad \mbox{or equivalently}\quad
\int(f_{i|j \gamma}-f_{i|\gamma}) f_{j|k\gamma}\,  \mathrm{d} y_j =0 .
\]

\begin{figure}[b]

\includegraphics{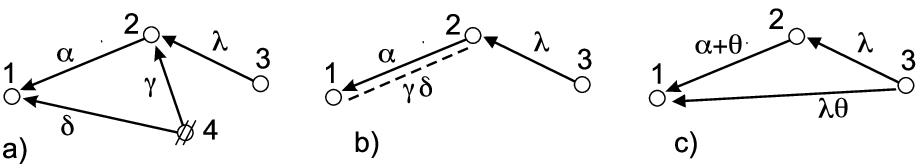}

\caption{\textup{(a)} Generating graph for Gaussian relations in standardized
variables, leading for variable $Y_4$ unobserved to \textup{(b)} the summary
graph and \textup{(c)} the maximal ancestral graph for the observed variables;
with the generating dependences attached to the arrows in \textup{(a)}, simple
correlations $
\rho_{12}=\alpha+\gamma\delta, \hspace*{.35em}\rho_{13}=\alpha
\lambda, \hspace*{.35em}
\rho_{23}=\lambda\mbox{ and } \theta=\gamma\delta/(1-\lambda^2)
$ are implied.}
\label{fig5}
\end{figure}

\subsection{Using summary graphs to detect distortions of generating
dependences}
In a MAG, the dependence corresponding to an $ik$-arrow may differ, without
any warning,
qualitatively from the generating dependence of $Y_i$ on $Y_k$
in $f_V$. In particular, it may change the sign but stay a strong
dependence. If this remained undetected, one would come to
qualitatively wrong conclusions when interpreting the parameters
measuring the conditional dependence of $Y_i$ on $Y_k$ in $f_{u|vC}$.

\begin{figure}[t]

\includegraphics{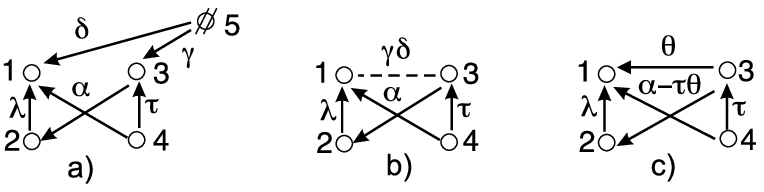}

\caption{\textup{(a)} Generating graph for linear relations in standardized
variables, leading for variable $Y_5$ unobserved to \textup{(b)} the summary
graph and \textup{(c)} the maximal ancestral graph for the observed variables;
with the generating dependences attached to the arrows in \textup{(a)} implied are:
$\theta=\gamma\delta/(1-\tau^2)$; generating dependence
$\lambda$ undistorted in both models to the graphs \textup{(b)}, \textup{(c)}; generating
dependence $\alpha$ preserved with~\textup{(b)}, distorted with \textup{(c)}.}
\label{fig6}
\end{figure}

The summary graph corresponding to a MAG detects, whether and for which
of the generating dependences, $i\fla k$,
having both of $i,k$ within $u$, such distortions can occur due
to direct or indirect confounding; see \citet{WerCox08} and Corollary~\ref{c4},
Lemma~\ref{l1} below. We illustrate here direct confounding with
Figure~\ref{fig5}
and indirect confounding with Figure~\ref{fig6}.

For a joint Gaussian distribution, the distortions are compactly
described in terms of regression coefficients for variables $Y_i$
standardized to have mean zero and variance one. For Figure~\ref{fig5}(a),
the generating equations are
%
\begin{equation}\label{eq1}
Y_{1}=\alpha Y_{2}+ \delta  Y_{4}+\varepsilon_{1} , \qquad Y_{2}=
\lambda Y_{3}+ \gamma Y_4+\varepsilon_{2 }, \qquad
Y_{3}=\varepsilon_{3}, \qquad Y_{4}=\varepsilon_{4}.
\end{equation}

With residuals having zero means and being
uncorrelated,
the equations of the summary graph model that result from \eqref{eq1}
for $Y_4$ unobserved have one pair of correlated residuals
\begin{eqnarray*}
&Y_{1}=\alpha Y_{2}+\eta_1 , \qquad Y_{2}= \lambda Y_{3}+\eta_2, \qquad
Y_{3}=\eta_{3},&
\\
&\eta_1=\delta  Y_{4}+\varepsilon_{1}, \qquad\eta_2= \gamma
Y_4+\varepsilon_{2 }, \qquad\eta_3=\varepsilon_3, \qquad
\operatorname{cov}(\eta
_1, \eta_2) = \gamma\delta.&
\end{eqnarray*}
The equation parameters of the standardized Gaussian associated with
the MAG of Figure~\ref{fig5}(c) are instead defined via
\[
E(Y_{1}|Y_{2}=y_2, Y_3=y_3), \qquad E(Y_{2}| Y_3=y_3),
\]
with all residuals in the recursive equations being uncorrelated.
The generating dependence $\alpha$ is retained in the summary graph model.

The parameter for the dependence of $Y_1$ on $Y_2$ in the MAG model,
expressed in terms of the generating parameters of Figure~\ref{fig5}(a), is
$\alpha+\gamma\delta/(1-\lambda^2)$. The summary
graph in Figure~\ref{fig5}(b) is a graphic representation of the simplest
type of an instrumental variable
model, used in econometrics [see \citet{Sargan58}] to separate a direct
confounding effect,
here $\gamma\delta$, from the dependence of interest, here $\alpha$.

In general, possible distortions due to direct confounding in
parameters of dependence in MAG models are recognized in the
corresponding summary graph by a double edge $i \matrix{\fla\vspace*{-8pt}\cr \dal\vspace*{1pt}} k$.
In the following example of Gaussian standardized variables, there is
no direct
confounding of the generating dependence $\alpha$ but there is
indirect confounding of $\alpha$ while $\lambda$ remains undistorted.

To simplify the figures, the coefficient attached to
$2\fla3$
is not displayed in any of the three graphs of Figure 6. The generating
graph in Figure~\ref{fig6}(a)  is directed and acyclic so that
the corresponding linear equations in standardized Gaussian variables,
defined implicitly by Figure~\ref{fig6}(a), have uncorrelated residuals.
The example is adapted from \citet{RobWas97}.
The summary graph in Figure~\ref{fig6}(b) shows with a dashed line the
induced association for pair $Y_1, Y_3$ that results by marginalising
$f_V$ over $Y_5$.

The equations of the summary graph model, obtained for $Y_5$
unobserved, have precisely one pair of correlated residuals,
$\operatorname{cov}(
\eta_1, \eta_3)=\gamma\delta$ and
\[
Y_{1}=\lambda Y_2+\alpha Y_{4}+\eta_1 , \qquad Y_{2}= \rho_{23}
Y_{3}+\eta_2, \qquad Y_{3}=\tau Y_4+\eta_{3}, \qquad Y_4=\eta_4.
\]
The summary graph model preserves both $\lambda$ and $\alpha$ as
equation parameters.

In the corresponding MAG model, represented by the graph in Figure
\ref{fig6}(c), the equation parameters associated with arrows present in
the graph are unconstrained linear least-squares regression coefficients.
These coefficients, expressed in terms of the generating parameters of
Figure~\ref{fig6}(a), are shown next to the arrows in Figure \ref{fig6}(c).
Thus, the generating coefficient $\lambda$ is preserved, while $\alpha
$ is changed into $\alpha-\tau\theta$, with $\theta=\gamma\delta
/(1-\break\tau^2)$.

Direct confounding of a generating dependence of $Y_i$ on $Y_k$ is
avoided in intervention studies, such as experiments and controlled
clinical trials,
by randomized allocation of individuals to the levels of $Y_k$, but
severe indirect confounding may occur nevertheless; see \citet{WerCox08}.

Then, the set of ancestors of node $i$ in $G^V_{\mathrm{par}}$ be
denoted by $\operatorname{anc}
_i$. Then, the set of ancestors of node $i$ in $G^{V\setminus
[C,M]}_{\mathrm{sum}}$
within $u$ is $c_i= u \cap\operatorname{anc}_i $ since
no additional ancestor of $i$ is ever generated within~$u$. Then, by
conditioning $Y_i$ on $Y_v$ and $Y_{c_i}$, one marginalises implicitly
over the nodes in set $m_i= \{ \{1, \ldots, i\}, \{u\cap\operatorname{pst}
_i\setminus c_i\}\}$ and indirect confounding
may result.

\begin{coro}[(Lack of confounding in measures of conditional
dependence)]\label{c3} A~generating dependence $i\fla k$ present in
$G^V_{\mathrm{par}}$ is
undistorted in the \textup{MAG} model in nodes $V\setminus\{C,M\}$: \textup{(1)}~by
direct confounding if in $G^V_{\mathrm{par}}$ there is no active
$ik$-path
relative to $\{C,M\}$ and \textup{(2)} by
indirect confounding if in $G_{\mathrm{sum}}^{V\setminus[C,M]}$
there is no
active $ik$-path relative to $\{c_i,m_i\}$.
\end{coro}

In distributions generated over $G_{\mathrm{par}}^V$, every active
path is
association-inducing, hence a~generating dependence will be confounded unless
the distortion is cancelled by other edge-inducing paths. When a
distortion is judged to be severe depends on the subject matter
context. To detect
indirect confounding,
we name $k$ a forefather of $i$ if it is an ancestor but not a parent
of $i$ and three dots indicate more edges and nodes of the same
type.

\begin{lemma}[(A graphical criterion {[\citet{WerCox08}]})]\label{l2}
For $i \fla k$ of $G_{\operatorname{par}}^V$, indirect confounding in
the absence of
direct confounding is generated in the {\rm MAG} model
by marginalising over $M=\{l>k, l+1, \dots, d_V\}$
if and only if in the corresponding summary graph $G^{V\setminus
[\varnothing, M]}_{\mathrm{sum}}$, which is without double edges,
associations for $Y_i, Y_k$ do not cancel that result by conditioning
on all ancestors of node~$i$, that is, from the following types of
collision $ik$-paths that have as inner nodes only forefathers of node $i$:
%
\begin{equation}
i \dal
\raisebox{-1pt}{\mbox{\includegraphics{309i01.eps}}}
\cdots
\raisebox{-1pt}{\mbox{\includegraphics{309i01}}}\dal
\raisebox{-1pt}{\mbox{\includegraphics{309i01}}}\dal k,
\qquad
i \dal
\raisebox{-1pt}{\mbox{\includegraphics{309i01.eps}}}
\cdots
\raisebox{-1pt}{\mbox{\includegraphics{309i01.eps}}}\dal
\raisebox{-1pt}{\mbox{\includegraphics{309i01}}}\fla k.
\end{equation}
\end{lemma}

An example of such a path of indirect confounding is given with Figure
\ref{fig6}(b) above, where for $1\fla4$, it is the path $1 \dal3 \fla4$.

In the following two sections, we give further preliminary results and
those proofs of new results for which we use more technical arguments.

\section{Further preliminary results}\label{sec2}
The edge matrix ${\mathcal{A}}$ of a parent graph
is a $d_V\times d_V$ unit upper-triangular
matrix, that is, a matrix with ones along the diagonal and zeros in the
lower triangular part, such that for $i<k$, element ${\mathcal
{A}}_{ik}$ of
${\mathcal{A}}$ satisfies
%
\begin{equation} \label{acal} {\mathcal{A}}_{ik}=1 \quad\mbox{if and
only if}\quad i
\fla k \mbox{ in } G_{\operatorname{par}}^V.
\end{equation}
Because of the triangular form of the edge matrix ${\mathcal{A}}$ of
$G_{\operatorname{par}
}^V$, a density $f_V$ generated over a given parent graph has also been
called a triangular system of densities.

\subsection{Linear triangular systems}

A linear triangular system is given by a set of recursive linear
equations for a mean-centered random vector variable $Y$ of dimension $
d_V \times1$ having $\operatorname{cov}(Y)=\Sigma$, that is, by
%
\begin{equation}\label{lsem2}
A Y = \varepsilon,
\end{equation}
where $A$ is a real-valued $d_V\times d_V$ unit upper-triangular
matrix, given by
\[
E_{\rm lin}( Y_i| Y_{i+1}=y_{i+1}, \dots, Y_{d_V}=y_{d_V})=-A_{i,
\operatorname{par}_i}y_{\operatorname{par}_i},
\]
and $E_{\rm lin}(\cdot)$ denotes a linear predictor.
The random vector $\varepsilon$ of residuals has zero mean and
$\operatorname{cov}
(\varepsilon)=\Delta$, a diagonal matrix. A Gaussian triangular
system of densities is generated if the distribution of each residual
$\varepsilon_i$ is Gaussian and the corresponding joint Gaussian
family varies fully if $\Delta_{ii}>0$ for all $i$.

The covariance and concentration matrix
of $Y$ are, respectively, using $(A^{-1})^{\mathrm{T}}=A^{-\rm T}$
%
\begin{equation} \label{covcon}
\Sigma= A^{-1} \Delta A^{-\rm T}, \qquad \Sigma^{-1}
= A^{\mathrm{T}}\Delta^{-1} A.
\end{equation}
Linear independences that constrain the equation \eqref{lsem2}
are
defined by zeros in the triangular decomposition, $(A, \Delta^{-1})$,
of the concentration matrix. For joint Gaussian distributions
\[
A_{ik}=0 \quad\iff\quad i\perp\!\!\!\hspace*{-1pt}\perp k|\operatorname{par}_i \mbox{ for }
k\in\operatorname{pst}_i\setminus\operatorname{par}_i.
\]

The edge matrix ${\mathcal{A}}$ of $G_{\operatorname{par}}^V$
coincides for Gaussian
triangular systems generated over $G_{\operatorname{par}}^V$ with the indicator
matrix of zeros in $A$, that is,
${\mathcal{A}}=\operatorname{In}[A] $,
where $\operatorname{In}[\cdot]$ changes every non-zero entry of a
matrix into a
one. Furthermore, since the parent graph in node set~$V$ is
edge-minimal for $f_V$, we have
\[
A_{ik}=0 \quad\iff\quad {\mathcal{A}}_{ik}=0.
\]

Edge matrices expressed in terms of components of a set of given
generating edge matrices are called induced. Simple examples of edge
matrices induced by ${\mathcal{A}}$ of \eqref{acal} are the overall
covariance and the overall concentration graph; see \citet{WerCox04}.
These two types of graphs have as induced edge matrices, respectively,
%
\begin{equation}\label{edcovcon}
{\mathcal{S}}_{VV}=\operatorname{In}[{\mathcal
{A}}^{-}{\mathcal{A}}^{- \rm T}] \quad\mbox{and}\quad {\mathcal{S}}^{VV}=\operatorname{In}[{\mathcal{A}}^{\rm T}
{\mathcal{A}}],
\end{equation}
where ${\mathcal{A}}^{-}$ has all ones of ${\mathcal{A}}$ and an additional
one in position $(i,k)$ if and only if $k$ is a~forefather of node $i$
in $G^V_{\operatorname{par}}$. In the graph with edge matrix ${\mathcal
{A}}^{-}$, every
forefather $k$ of~$i$
is turned into a parent, that is, $i\fla k$ is inserted.

By writing the two matrix products in \eqref{edcovcon} explicitly, one sees
that for an uncoupled node pair $i,k$ in the parent graph, there is an
additional edge in the induced concentration graph of $Y_V$ if and only
if the pair has a common offspring in $G_{\operatorname{par}}^V$. With
a zero in
position $i,k$ of ${\mathcal{A}}^{-}$, there is an additional $ik$-edge in the
induced covariance graph if and only if an uncoupled pair has a common
parent in the directed graph with edge ma\-trix~${\mathcal{A}}^{-}$.\looseness=1

Both of these induced matrices are symmetric. The covariance and the
concentration matrix, implied by a linear triangular system and given in
\eqref{covcon},
contain all zeros present in the corresponding induced edge matrices,
but possibly more. This happens for $(i,k)$ whenever the associations
that are induced for
$Y_i,Y_k$ by several edge-inducing $ik$-paths cancel precisely. For
such particular parametric constellations in Gaussian distributions
generated over parent graphs, see \citet{WerCox98}.
In data analyses, near cancellations are encountered frequently.

By contrast, the induced edge matrices capture consequences of the
generating independence structure. They contain structural zeros. These are zeros that occur for all permissible
parametrisations, or, expressed differently, that occur for each member
of a~family $f_V$ generated over a given $G_{\operatorname{par}}^V$.

For distributions generated over parent graphs, a zero in position
$(i,k)$ of ${\mathcal{S}}_{VV}$ and of
${\mathcal{S}}^{VV}$ means, respectively, that
%
\begin{equation}\label{overallg}
 i\perp\!\!\!\hspace*{-1pt}\perp k, \qquad i\perp\!\!\!\hspace*{-1pt}\perp
k|V\setminus\{i,k\}
\end{equation}
is implied by $G_{\operatorname{par}}^V$.
Thus, in contrast to the global Markov property,
the induced graphs answer all queries concerning sets of these two
types of independence statements at once.

More complex induced edge matrices arise, for instance, in regression graphs and
in summary graphs derived from ${\mathcal{A}}$.
For transformations of linear systems, we use the operator called
partial inversion, which is introduced next; for proofs and discussions
see \citet{WerWieCox06}, \citet{MarWer09}, \citet{WieWer10}.

\vspace*{3pt}
\subsection{Partial inversion}
\vspace*{3pt}

Let $F$ be a square matrix of dimension $d_V$ with principal
submatrices that are all invertible. This holds for every $A$ of \eqref
{lsem2} and for every covariance matrix of a Gaussian distribution that
varies fully, so that $Y$ has no degenerate component.

For any subset $a$ of $V$ and $b=V\setminus a$, by applying
partial inversion to
the linear equations $FY=\eta$, say, these are modified into
%
\begin{equation}\label{eqkey}
{\operatorname{inv}}_{a} F \pmatrix{
\eta_{a}\cr
Y_{b}
} = \pmatrix{
Y_{a}\cr
\eta_{b}
}.
\end{equation}
By applying partial inversion to $b$ of $V$ in equation \eqref{eqkey},
one obtains $Y=F^{-1}\eta$. Thus, full inversion is decomposed
into two steps of partial inversion.

Partial inversion extends the sweep operator for symmetric, invertible
matrices to non-symmetric matrices $F$
%
\begin{equation}
\label{defpinv}
\operatorname{inv}_{a}F=\pmatrix{
F_{aa}^{-1}& -F_{aa}^{-1}{F}_{ab}\cr
F_{ba}F_{aa}^{-1}&F_{bb.a}
}
\qquad \mbox{with } F_{bb.a}=F_{bb}-
F_{ba}F_{aa}^{-1}
F_{ab}.
\end{equation}

\begin{lemma}[(Some properties of partial
inversion {[Wermuth, Wiedenbeck and Cox \citeyear{WerWieCox06}]})]\label{l3} Partial inversion is commutative, can be undone and is exchangeable
with selecting a submatrix. For $V$ partitioned as $V=\{a,b,c,d\}$:
\begin{longlist}[(1)]
\item[(1)]$\operatorname{inv}_{a} \operatorname
{inv}_{b}F=\operatorname{inv}_{b} \operatorname{inv}_{a}F$,

\item[(2)]$\operatorname{inv}_{ab} \operatorname
{inv}_{bc} F=\operatorname{inv}_{ac} F,$

\item[(3)]$[\operatorname{inv}_{a}
F]_{J,J}=\operatorname{inv}_{a}F_{JJ} \mbox{ for } J=\{
a,b\}$.
\end{longlist}
\end{lemma}

In contrast, the sweep operator cannot be undone; see \citet{Dem72}.
Example~\ref{e1} shows how the triangular equations in \eqref{lsem2} are
modified by partial inversion on~$a$, where $a$~consists of the first
$d_a$ components of $Y$. Instead of the full recursive order
$V=(1,\ldots, d_V)$ with uncorrelated residuals, a block-recursive order
$V=(a,b)$ results, where residuals within $a$ are correlated, but
uncorrelated with the unchanged residuals within $b$.

\begin{exa}[(Partial inversion applied to a linear triangular
system \textup{\eqref{lsem2}} with an order-respecting split of $\bolds{V}$)]\label{e1} For $a=\{
1, \ldots, d_a\}$,
$b=\{d_{a} +1, \dots, d_V\}$
\[
\operatorname{inv}_{a}A=\pmatrix{
A_{aa}^{-1}& -A_{aa}^{-1}A_{ab}\cr
0&A_{bb}
}
\qquad \mbox{gives with } Y_a =- A_{aa}^{-1}A_{ab}Y_b+A_{aa}^{-1}\varepsilon_a,
\]
the implied form of linear least-squares regression of $Y_a$ on $Y_b$, where
\[
E_{  \rm lin}(Y_a|Y_b=y_b)=\Pi_{a|b}y_b, \qquad
Y_{a|b}=Y_a-\Pi_{a|b}Y_b, \qquad\operatorname{cov}
(Y_{a|b})=\Sigma_{aa|b}
\]
and
\[
\Pi_{a|b}=- A_{aa}^{-1}A_{ab},
\qquad\Sigma
_{aa|b}=A_{aa}^{-1}\Delta_{aa}A^{-\rm T}_{aa}.
\]
\end{exa}

Example~\ref{e2} shows how the triangular equations contained in \eqref{lsem2} are
modified by partial inversion on $b$, where $V=(a,b,c)$ so that $b$
consists of
intermediate components of $Y$. To use the matrix formulation in \eqref
{defpinv} directly, one sets $b:=(a,c)$, $a:=b$ and leaves components
within $a$ and $b$ unchanged to obtain $\tilde{A}$, which is not
block-triangular in $(a,b).$
After partial inversion of $\tilde{A}$ on $a$, the original order is
restored for the results presented in Example~\ref{e2}.

\begin{exa}[(Partial inversion applied to a linear triangular
system \textup{\eqref{lsem2}} for an order-respecting partitioning
$\bolds{V=(a,b,c)}$)]\label{e2} With $a=\{1, \ldots, d_a\}$, $b=\{d_{a} +1, \dots,
(d_a+d_b)\}$ and $c=\{(d_a+d_b)+1, \dots, d_V\}$,
\[
\operatorname{inv}_{b}A=\pmatrix{
A_{aa}& A_{ab} A_{bb}^{-1}&
A_{ac.b}\cr
0&A_{bb}^{-1} & -A_{bb}^{-1} A_{bc}
\cr
0& 0 & A_{cc}
} \qquad\mbox{gives } Y_a =- A_{aa}^{-1}A_{ac.b}Y_c+\eta_a,
\]
the implied form of the linear least-squares regression of $Y_a$ on
$Y_c$, with
\[
\eta_a=A_{aa}^{-1}\varepsilon_a+\Pi_{a|b.c}
A_{bb}^{-1}  \varepsilon_b, \qquad\Pi
_{a|bc}=(\Pi_{a|b.c}, \Pi_{a|c.b})=- A_{aa}^{-1}(A_{ab} ,
A_{ac}).
\]
For $\Pi_{a|c}$, a special form of Cochran's recursive
definition of regression coefficients results, see also \citet{WerCox04},
\[
\Pi_{a|c}=\Pi_{a|c.b}+\Pi
_{a|b.c}\Pi_{b|c}=-A_{aa}^{-1}(A_{ac}- A_{ab}A_{bb}^{-1}A_{bc}
)=-A_{aa}^{-1}A_{ac.b}.
\]
For $\operatorname{cov}(Y_{a|c})$, Anderson's recursive definition of
covariance
matrices results:
\[
\Sigma_{aa|c}=A_{aa}^{-1}\Delta_{aa} A_{aa}^{-\rm
T}+\Pi
_{a|b.c}(A_{bb}^{-1}\Delta_{bb}A_{bb}^{-\rm T})\Pi
_{a|b.c}^{\mathrm{T}}=\Sigma_{aa|bc}+\Sigma_{ab|c}\Sigma
_{bb|c}^{-1} \Sigma_{ba|c}.
\]
\end{exa}

For $b,c$, the result in Example~\ref{e2} is as in Example~\ref{e1}. For $Y_a$, the
original recursive regressions given $Y_b,Y_c$ are modified into
recursive regressions given only $Y_c$. The residuals between $Y_a,Y_b$
are correlated since $\operatorname{cov}(Y_{a|c}, Y_{b|c})=\Sigma
_{ab|c}$ but
remain uncorrelated from those in $c$. In the modified equations, $Y_b$
can be removed without affecting any of the other remaining relations.

For a more detailed discussion of the three different types of
recursion relations of linear association measures due to Cochran,
Anderson and Dempster,
see Wiedenbeck and Wermuth \citeyear{WieWer10}.

For Example~\ref{e3}, one starts with equation \eqref{lsem2} premultiplied
by $A^{\rm T} \Delta^{-1}$ and obtains linear equations in which the
equation parameter matrix, $\Sigma^{-1}$, coincides with the
covariance matrix of the residuals, that is, one starts with
%
\begin{equation}\label{lconcm}
 \Sigma^{-1} Y=A^{\rm T} \Delta^{-1}\varepsilon.
\end{equation}

\begin{exa}[(Partial inversion with any split of $\bolds{V}$ applied to
$\bolds{\Sigma^{-1}}$)]\label{e3} The covariance matrix $\Sigma$ and the concentration
matrix $\Sigma^{-1}$ of $Y$ are written, partitioned according to
$(a,b)$ for $a$ any subset of $V$, as
\[
\Sigma=
\pmatrix{ \Sigma_{aa} & \Sigma_{ab} \cr
\cdot & \Sigma_{bb}
}
, \qquad
\Sigma^{-1}=
\pmatrix{ \Sigma^{aa} & \Sigma^{ab} \cr
\cdot & \Sigma^{bb}
}
,
\]
where the $\cdot$ notation indicates symmetric entries.
Partial inversion of $\Sigma^{-1}$ on $a$ leads to three distinct
components, $\Pi_{a|b}$, the population coefficient matrix of $Y_b$ in
linear least-squares regression of $Y_a$ on $Y_b$;
the covariance matrix $\Sigma_{aa|b}$ of $Y_{a|b}$;
and the marginal concentration matrix $\Sigma^{bb.a}$ of $Y_b$
%
\begin{equation}
\label{S1}
\operatorname{inv}_a \Sigma^{-1} =
\pmatrix{\Sigma_{aa|b} & \Pi_{a|b} \cr
\sim& \Sigma^{bb.a}
},
\end{equation}
where the $ {\scriptstyle\sim}$ notation denotes entries that are
symmetric except for the sign.

Since \eqref{eqkey} and \eqref{defpinv} give $\operatorname{inv}_a
\Sigma
^{-1}=\operatorname{inv}_b  \Sigma$ directly, several well-known
dual expressions
for the three submatrices in \eqref{S1} result:
\[
\pmatrix{(\Sigma^{aa})^{-1} & -(\Sigma^{aa})^{-1}\Sigma^{ab}\cr
\sim& \Sigma^{bb}-\Sigma^{ba}(\Sigma^{aa})^{-1}\Sigma^{ab}
}
=
\pmatrix{
\Sigma_{aa} -\Sigma_{ab}\Sigma
_{bb}^{-1}\Sigma_{ba} & \Sigma_{ab}\Sigma_{bb}^{-1}
\cr
\sim& \Sigma_{bb}^{-1}
},
\]
where the explicit form of $\Sigma^{-1}_{bb}=\Sigma^{bb.a}$ is Dempster's recursive definition of
concentration matrices.
\end{exa}

A more complex key result is that, for any block-triangular system of
linear equations for $Y$, with equation parameter matrix $H$ and with
possibly correlated residuals obtained from $W=\operatorname
{cov}(HY)$, the implied
form of $\operatorname{inv}_a \Sigma^{-1}$ can be expressed in terms
of partially
inverted matrices $H$ and $W$.

Linear equations in a mean-centered vector variable $Y$ are
block-triangular in two ordered blocks $ (a,b)$ with
a positive-definite $\Sigma^{-1}=H^{\mathrm{T}} W^{-1}H $ if
%
\begin{equation}\label{blockdiag}
 H Y=\eta, \qquad\mbox{with } H_{ba}=0, \qquad E(\eta)=0,
\qquad\operatorname{cov}(\eta)=W \mbox{ positive-definite}.
\end{equation}
For $ K=\operatorname{inv}_a H$ and $Q=\operatorname{inv}_b W, $
direct computations give
%
\begin{equation}\label{invprod1}
\operatorname{inv}_{a} (H^{\mathrm{T}} W^{-1}H )\!=\!
\pmatrix{K_{aa}Q_{aa}K_{aa}^{\mathrm{T}}&
K_{ab}+K_{aa}Q_{ab}K_{bb}
\cr
\sim&H_{bb}^{\mathrm{T}} Q_{bb}H_{bb}
}.
\end{equation}
A simple special case is the triangular linear system \eqref{lsem2}.
Example~\ref{e4} shows how regressions in blocks $(a,b)$ result from it.

\begin{exa}\label{e4} For \eqref{blockdiag} with $H=A$ of \eqref{lsem2},
$W=\Delta$ diagonal and $a=1, \ldots, d_a$,
\[
\operatorname{inv}_{a} (H^{\mathrm{T}}\Delta^{-1}H )=
\pmatrix{\Sigma_{aa|b}&    \Pi_{a|b}\cr
\sim&   \Sigma_{bb}^{-1}
}
=
\pmatrix{K_{aa}\Delta_{aa}K_{aa}^{\mathrm{T}}&   K_{ab}\cr
\sim&A_{bb}^{\mathrm{T}}\Delta_{bb}^{-1}A_{bb}
}
.
\]
\end{exa}

Other special cases of linear block-triangular systems \eqref{blockdiag}
are Gaussian summary graph models; see Section~\ref{sec3}.

\subsection{Partial closure}

Let ${\mathcal{F}}$ be a binary edge matrix for node set $V=\{1,
\ldots, d_V\}
$ associated with $F$.
The operator called partial closure transforms ${\mathcal{F}}$ into
$\operatorname{zer}
_{a}{\mathcal{F}}$ so that in the corresponding
graph $a$-line paths of a special type become closed. For instance,
applied to ${\mathcal{A}}$, every $a$-line ancestor of node $i$ is
turned into
a parent of $i$ and, applied to the edge matrix of an undirected graph,
such as ${\mathcal{S}}^{VV}$, every $a$-line path is closed. Zeros in
the new
binary matrix $\operatorname{zer}_{a}{\mathcal{F}}$ are
the structural zeros that remain of $\operatorname{inv}_{a}F$.

In matrix form, with $n-1=d_a$ and ${\mathcal{I}}_{aa}$ a $d_a \times d_a$
identity matrix,
%
\begin{eqnarray}
\label{defzer}
\operatorname{zer}_a {{\mathcal{F}}} &=&\operatorname{In}\biggl[\pmatrix{
{\mathcal{F}}_{aa}^{-} & {\mathcal
{F}}^{-}_{aa}{\mathcal{F}}_{ab}\cr
{\mathcal{F}}_{ba}{\mathcal{F}}^{-}_{aa}&{\mathcal{F}}_{bb.a}
}\biggr] \qquad\mbox{ with }{\mathcal{F}}_{bb.a}=\operatorname
{In}[{\mathcal{F}}_{bb} + {\mathcal{F}}_{ba}
{\mathcal{F}}^{-}_{aa}{\mathcal{F}}_{ab}],\quad
\\\label{eqinvm}
{\mathcal{F}}_{aa}^{-}&=&\operatorname{In}[(n  {\mathcal{I}}_{aa} -
{\mathcal{F}}_{aa})^{-1}].
\end{eqnarray}
The inverse in \eqref{eqinvm} assures non-negative entries in $
{\mathcal{F}}
_{aa}^{-}$ and is a type of regularization; see \citet{Tikh63}. It
generalizes limits of scalar geometric series; see \citet{Neum1884},
page~29.

\begin{lemma}[(Some properties of partial
closure {[\citet{WerWieCox06}]})]\label{l4}
Partial closure is commutative, cannot be undone and is exchangeable
with selecting a submatrix. For $V$ partitioned as $V=\{a,b,c,d\}$:
\begin{enumerate}[(1)]
\item[(1)] $\operatorname{zer}_{a} \operatorname
{zer}_{b}F=\operatorname{zer}_{b} \operatorname{zer}_{a}F$,

\item[(2)] $\operatorname{zer}_{ab} \operatorname
{zer}_{bc} F=\operatorname{zer}_{abc} F,$

\item[(3)] $[\operatorname{zer}_{a}
F]_{J,J}=\operatorname{zer}_{a}F_{JJ} \mbox{ for } J=\{
a,b\}$.
\end{enumerate}
\end{lemma}

Given Gaussian parameter matrix components after partial inversion,
such as in equation~\eqref{invprod1}, the corresponding induced edge
matrices are obtained using Lemma~\ref{l5}, provided each component matrix
belongs to the model of the starting graph and the expressions
are minimal, that is, condensed in such a way that they do not contain
any parameter matrices that cancel, as, for instance, $A_{aa}^{\hspace
*{.35em}}A_{aa}^{-1}$ would.

\begin{lemma}[(Edges induced by a starting
graph obtained with minimal matrix expressions of Gaussian parameter
matrices {[\citet{MarWer09}]})]\label{l5} Edge matrices replace corresponding parameter matrices
after:
\begin{enumerate}[(1)]
\item[(1)] changing each negative sign to a positive sign,

\item[(2)] replacing in the resulting expressions each diagonal matrix by an
identity matrix or deleting it if it arises within a matrix
product,
and then applying the indicator function.
\end{enumerate}
\end{lemma}

For instance, the matrix formulation of partial inversion in \eqref
{defzer} can be viewed as arising from \eqref{defpinv} by use of Lemma~\ref{l5}.

\setcounter{exa}{0}
\begin{exa}[(Continued)]
Let ${\mathcal{K}}_{aa}={\mathcal{A}}_{aa}^{-}$ and ${\mathcal
{K}}_{ab}={\mathcal{A}}_{aa}^{-}{\mathcal{A}}_{ab}$.
After partial closure in $G^V_{\operatorname{par}}$ on $a$, there are
two induced edge
matrix components. For directed edges, it is $\operatorname{zer}_{a}
{\mathcal{A}}$, and
for undirected dashed line edges, it is ${\mathcal{S}}_{aa|b} $\vspace*{-1pt}
\[
\operatorname{zer}_{a} {\mathcal{A}}=\operatorname{In}\biggl[\pmatrix{
{\mathcal{K}}_{aa}& {\mathcal{K}}_{ab}\cr
0&{\mathcal{A}}_{bb}
}\biggr],
\qquad{\mathcal{P}}_{a|b}=\operatorname
{In}[{\mathcal{K}}_{ab}], \qquad{\mathcal
{S}}_{aa|b}=\operatorname{In}
[{\mathcal{K}}_{aa} {\mathcal{K}}_{aa}^{\mathrm{T}}].\vspace*{-1pt}
\]
\end{exa}

The induced graph of two components is a regression graph.

\begin{exa}[(Continued)]
By marginalising over the intermediate node set $b$ of $V=(a,b,c)$ in
$G^V_{\operatorname{par}}$, a directed acyclic graph results. The
induced Gaussian
parameter and edge matrices are, for $N=V\setminus b$, respectively,\vspace*{-1pt}
\[
[\operatorname{inv}_{b} A ]_{N,N}=
\pmatrix{ A_{aa}& A_{ac.b}\cr
0 & A_{cc}
}
, \qquad[\operatorname{zer}_{b} {\mathcal
{A}}]_{N,N}=\operatorname{In}\biggl[
\pmatrix{ {\mathcal{A}}_{aa}& {\mathcal
{A}}_{ac.b}\cr
0 & {\mathcal{A}}_{cc}
}
\biggr].\vspace*{-1pt}
\]
\end{exa}

\begin{exa}[(Continued)] A
concentration graph has for joint Gaussian distributions $\Sigma^{-1}$ as
the parameter matrix and ${\mathcal{S}}^{VV}$ as the edge matrix. By partial
closure on $a$ of ${\mathcal{S}}^{VV}$ given any split $V=\{a,b\}$, every
$a$-line path is closed.
Three edge matrix parts result:
${\mathcal{S}}_{aa|b}, {\mathcal{P}}_{a|b}$ and ${\mathcal
{S}}^{bb.a}$. They give the
structural zeros in the corresponding parameter matrices $ \Sigma
_{aa|b}, \Pi_{a|b}$ and $\Sigma^{bb.a}$. In general, the edge matrix
${\mathcal{S}}^{bb.a}$ is for the marginal concentration graph of $Y_b$.
\end{exa}

When the generating graph is $G^V_{\operatorname{par}}$, then a
concentration graph is
induced for the node set that contains ancestors of $C$ outside $C$. In
Example~\ref{e4}, the three components of $\operatorname{inv}_a \Sigma^{VV}$
are directly
expressed in terms of the triangular decomposition
$(A, \Delta^{-1})$.

\begin{exa}[(Continued)]
For the order-respecting split, $V=(a,b)$, and ${\mathcal
{K}}_{aa}={\mathcal{A}}
_{aa}^{-}$ and ${\mathcal{K}}_{ab}={\mathcal{A}}_{aa}^{-}{\mathcal
{A}}_{ab}$, a parent graph
$G^V_{\operatorname{par}}$ induces a regression graph for $f_{a|b}$ and
$f_{b}$ with
the following three edge matrix components\vspace*{-1pt}
%
\begin{equation}\label{pkey}
\pmatrix{ {\mathcal{S}}_{aa|b}& {\mathcal{P}}_{a|b}\cr
\cdot &{\mathcal{S}}^{bb.a}
}
=\operatorname{In}\biggl[
\pmatrix{{\mathcal{K}}_{aa}{\mathcal{K}}_{aa}^{\mathrm{T}}&   {\mathcal
{K}}_{ab}\cr
\cdot &{\mathcal{A}}_{bb}^{\mathrm{T}}{\mathcal{A}}_{bb}
}
\biggr] .\vspace*{-1pt}
\end{equation}
\end{exa}

The result combines the one in \eqref{edcovcon} in slightly modified
form with the above continuation of Example~\ref{e1} by considering the consequences
of a given parent graph for the distributions of $Y_a$ given $Y_b$ and
of $Y_b$.\vadjust{\goodbreak}

For the more complex generating graphs connected with block-triangular
linear systems \eqref{blockdiag} and given edge matrices ${\mathcal{H}},
{\mathcal{W}}$,
the three edge matrix components
in the induced regression graph of just two components are with
\begin{eqnarray}\label{zerprod}
{\mathcal{K}}&=&\operatorname{zer}_a {\mathcal{H}}, \qquad{\mathcal{Q}}=\operatorname{zer}_b
{\mathcal{W}},\nonumber\\
\pmatrix{ {\mathcal{S}}_{aa|b}& {\mathcal{P}}_{a|b}\cr \cdot
&{\mathcal{S}}^{bb.a}
}
&=&\operatorname{In}\biggl[
\pmatrix{{\mathcal{K}}_{aa}{\mathcal{Q}}_{aa}{\mathcal
{K}}_{aa}^{\mathrm{T}}&   {\mathcal{K}}
_{ab}+{\mathcal{K}}_{aa}{\mathcal{Q}}_{ab}{\mathcal{K}}_{bb}
\cr
\cdot &{\mathcal{H}}_{bb}^{\mathrm{T}}{\mathcal{Q}}_{bb}{\mathcal{H}}_{bb}
}
\biggr] .
\end{eqnarray}
From \eqref{zerprod}
for $a=\{\alpha, \delta\}$, the edge matrices induced by
$G^V_{\operatorname{par}}$
for $f_{\alpha|b} $ are
\[
{\mathcal{S}}_{\alpha\alpha|b}=[ {\mathcal{S}}_{aa|b}]_{\alpha,
\alpha}, \qquad
{\mathcal{P}}_{\alpha|b}=[{\mathcal{P}}_{a|b}]_{\alpha, b},
\]
and with a split of $b$ as $\{\beta, \gamma\}$, the edge matrix
induced for $f_{\beta|\gamma}$ and for the dependence of $Y_{\alpha
|\gamma}$ given $Y_{\beta|\gamma}$ are
\[
{\mathcal{S}}^{\beta\beta.a}=[{\mathcal{S}}^{bb.a}]_{\beta, \beta
}\quad \mbox{and}\quad
{\mathcal{P}}_{\alpha|\beta.\gamma}=[{\mathcal{P}}_{a|b}]_{\alpha
, \beta}.
\]

In general, the induced graphs of \eqref{pkey} or \eqref{zerprod}
with dashed lines for $ {\mathcal{S}}_{aa|b}$, arrows for ${\mathcal
{P}}_{a|b}$ and
full lines
for ${\mathcal{S}}^{bb.a} $ will not be independence-preserving
graphs. In
both graphs, the global Markov property of Lemma~\ref{l1} implies the meaning of
a missing $ik$-edge as
%
\begin{equation} \label{indMRC}
i \perp\!\!\!\hspace*{-1pt}\perp k| b \mbox{ in } {\mathcal{S}}_{aa|b}, \qquad
i \perp\!\!\!\hspace*{-1pt}\perp k|b\setminus k\mbox{ in } {\mathcal{P}}_{a|b} ,
\qquad
i\perp\!\!\!\hspace*{-1pt}\perp k| b\setminus\{i,k\}\mbox{ in }
{\mathcal{S}}^{bb.a}.\
\end{equation}

Whenever every edge-inducing path is association-inducing, conditional
dependences correspond to edges present in the graph in the resulting families of densities of $Y_{a|b}$, $Y_b$
and also in a given member of the family unless associations cancel that are due to several
edge-inducing paths.

\section{Summary graphs and associated models}\label{sec3}

\subsection{Gaussian summary graph models}

Starting from a Gaussian triangular system \eqref{lsem2} generated
over a parent graph in node set $V$, marginalising over $M$ and
conditioning on $C$ gives a linear system of equations for $Y_{N|C}$
for $N=(u,v)=V\setminus\{C,M\}$ of the following form, where for the
equations in the ancestors $v$ of $C$ that are outside of $C$, the
equation parameter matrix and the covariance matrix coincide with a
concentration matrix, as in \eqref{lconcm}.

\begin{defn}[(Gaussian summary graph model)] \label{def2}
A Gaussian summary graph model is a system of equations $HY_{N|C}=\eta
$ that is block-triangular and orthogonal in $(u,v)$ with
%
\begin{equation} \label{linsumm}
\pmatrix{ H_{uu} & H_{uv}\cr
0& \Sigma^{vv.uM}
}
\pmatrix{ Y_{u|C} \cr Y_{v|C}
}
=
\pmatrix{ \eta_{u}\cr \zeta_v
}
, \qquad
\operatorname{cov}
\pmatrix{ \eta_{u}\cr \zeta_v
}
=
\pmatrix{ W_{uu} & 0 \cr
\cdot& \Sigma^{vv.uM}
}
,
\end{equation}
where $H_{uu}$ is unit upper-triangular, $W_{uu}$ and $\Sigma
_{vv|C}^{-1}=\Sigma^{vv.uM}
$ are symmetric and each of $\eta_u$ and $\zeta_v$ have freely
varying joint Gaussian distributions.
The independence structure is given by a summary graph in node set $N$;
see Definition~\ref{d6} and Section~\ref{sec3.2} below.
\end{defn}

For $Y_{v|C}$, equation \eqref{linsumm} specifies a Gaussian
concentration graph model. These models had been studied under the name
of covariance selection by Dempster (\citeyear{Dem72}); see also \citet
{SpeedKiv86}. For each member of the this family of models, the
likelihood function has a unique maximum.

With $W_{uv}=0$, the residuals of $Y_{u|C}$ and $Y_{v|C}$ are
uncorrelated, therefore the
system of equation~\eqref{linsumm} is said to be orthogonal in
$(u,v)$. Because of this orthogonality,
$\Pi_{u|v.C}=-H_{uu}^{-1}H_{uv}$ is the population least-squares
regression coefficient matrix in linear regression of $Y_{u|C}$ on
$Y_{v|C}$; see Example~\ref{e1} above. In econometrics, the equation in
$Y_{u|C}$ resulting by premultiplication with $H_{uu}^{-1}$ from the
first equation of \eqref{linsumm} is called the reduced form.

The equation in $Y_{u|C}$ of \eqref{linsumm} can equivalently be
written as a recursive system in endogenous variables
$Y_{u|vC}=Y_{u|C}-\Pi_{u|v.C} Y_{v|C:}$\vspace*{-1pt}
%
\begin{equation} \label{recreg} H_{uu} Y_{u|vC}=\eta_u \qquad\mbox{with }
\operatorname{cov}(\eta_u) =W_{uu},\vspace*{-1pt}
\end{equation}
where the equation parameter matrix $H_{uu}$ is, as in the linear
triangular system \eqref{lsem2}, of unit upper-triangular form, but
some of the residuals $\eta_u$ are correlated. For estimation, one
speaks in econometrics of the endogeneity problem; see Drton, Eichler and Richard\-son
\citeyear{Drton09b} for a recent discussion.

Identification is an issue for estimating the equation parameters
$H_{uu}$ in \eqref{recreg}. No necessary and sufficient condition is
known yet;
see \citet{KangTian09}. One general sufficient condition is the
absence of any double edge in the summary graph; see Brito and Pearl \citeyear{BriPea02}.
This says that
for any pair $i,k$ within $u$, either $H_{ik}=0$, or $W_{ik}=0$, or
both hold.

However, some models with double edges in the $G^N_{\mathrm{sum}}$
correspond to
identified instrumental variable models; see the above example to
Figure~\ref{fig5}(b). For the identifiability of latent variable models,
which arise here via larger
hypothesized generating processes, the notion of completeness is again
relevant; see \citet{SanMMou07}.

\subsection{Generating $G^{V\setminus[C,M]}_{\mathrm{sum}}$ from
$G^{V}_{\operatorname{par}}$}\label{sec3.2}

The summary graph $G^{V\setminus[C,M]}_{\mathrm{sum}}$ has four
edge matrix components.
With ${\mathcal{S}}^{vv.uM}$ a concentration graph results in node set
$v$, with
${\mathcal{H}}_{uu}$ a directed acyclic graph within  $u$, with ${\mathcal
{W}}_{uu}$ a
covariance graph of the residuals $ \eta_{u}$ and with ${\mathcal
{H}}_{uv}$ a
bipartite graph for dependence of $Y_{u|C}$ on $Y_{v|C}$.

Starting from a Gaussian triangular system in \eqref{lsem2} with
parent graph $G^{V}_{\operatorname{par}}$,
the choice of any conditioning set $C$ leads to an ordered split $V
=(O,R)$, where we
think of $R=\{C, F\}$ as the nodes to the right of $O$; see equation
\eqref{OSmat}. Every node in
$F$ is an ancestor of a node in $C$ outside $C$, so that we call $F$
the set of foster nodes of $C$.
No node in $O$ has a descendant in $R$ so that $O$ is said to contain
the outsiders of $R$.
Equations, orthogonal and block-triangular in $(O,R)$, are in unchanged order\vspace*{-1pt}
%
\begin{equation} \label{OSmat}
\pmatrix{ A_{OO}& A_{OR}\cr
0 & A_{RR}
}
\pmatrix{ Y_{O} \cr Y_{R}
}
=
\pmatrix{ \varepsilon_O \cr \varepsilon_R
}
.\vspace*{-1pt}
\end{equation}
After conditioning on $Y_C$ and marginalising over $Y_{M}$, the
resulting system
preserves block-triangularity and orthogonality with $u \subseteq O$,
$v \subseteq F$.

\begin{prop}[(Linear equations obtained from
$\bolds{AY=\varepsilon}$ after conditioning on $\bolds{Y_C}$ and marginalising over $\bolds{Y_M}$)]\label{p4}
Given a Gaussian triangular system \eqref{lsem2} generated over
$G^{V}_{\operatorname{par}}$,\vadjust{\goodbreak} conditioning set $C$, marginalising set
$M=(p,q)$ with
\[
p =O\setminus u, \qquad q =F\setminus v,
\]
and partially inverted parameter matrices arranged in the appropriate order,
\[
D=\operatorname{inv}_{p}  \tilde{A}, \qquad\operatorname
{inv}_q \tilde{\Sigma}^{FF.O}=
\pmatrix{\Sigma_{qq|vC} & \Pi_{q|v.C} \cr
\sim& \Sigma^{vv.qO}
}
,
\]
the induced linear equation \eqref{linsumm} in $Y_{N|C}$ have
equation parameters
%
\begin{equation} \label{HuN} H_{uu}=D_{uu}, \qquad
H_{uv}=D_{uv}+D_{uq}\Pi_{q|v.C}, \qquad\Sigma^{vv.uM}
\end{equation}
and covariance matrices
%
\begin{equation} \label{Wuu} W_{uu}=(\Delta_{uu}+D_{up}\Delta
_{pp}D_{up}^{\mathrm{T}})+(D_{uq}\Sigma_{qq|vC}D_{uq}^{\mathrm{T}}), \qquad\Sigma
^{vv.uM}.
\end{equation}
\end{prop}

\begin{pf}
Equation \eqref{OSmat} in $Y\!$ are first modified into equations for
$Y_{O|C}$ and $Y_{F|C}$. As for Example~\ref{e3} above, one takes $\zeta
_{R}=A_{RR}
\Delta_{RR}^{-1}\varepsilon_R $. After noting that
\[
\Sigma_{FF|C}^{-1}=[\Sigma^{RR.O}]_{F,F}=\Sigma^{FF.O}
\]
and by the orthogonality in $(O,R) $, these equations can be written as
\[
A_{OO}Y_{O|C}+A_{OF}Y_{F|C}=\varepsilon_O, \qquad \Sigma^{FF.O}
Y_{F|C}=\zeta_{F}.
\]
Partial inversion on $M=(p,q) $ gives, after appropriate ordering,
%
\begin{equation}\label{pinvpq}\operatorname{inv}_M
\pmatrix{\tilde{A}_{OO} & \tilde{A}_{OF}\cr
0& \tilde{\Sigma}^{FF.O}
}
\pmatrix{ \varepsilon_p \cr Y_{u|C} \cr \zeta'_{q} \cr Y_{v|C}
}
=
\pmatrix{ Y_{p|C}\cr \varepsilon_u\cr Y_{q|C}\cr \zeta'_{v}
}
,
\end{equation}
where, after deleting the equations in $Y_{M|C}$, the uncorrelated
residuals are
\[
\eta_{u}=(\varepsilon_u -D_{up}\varepsilon_p)-D_{uq}\Sigma
_{qq|vC}\zeta_{q},\qquad \zeta_{v}=\zeta
'_{v} + \Pi_{q|v.C}^{\mathrm{T}}
\zeta'_{q}.
\]
Thus, the equation parameter matrices of \eqref{HuN} and the
covariance matrices of \eqref{Wuu} result, where $\Sigma
_{vv|C}^{-1}=\Sigma^{vv.qO}=\Sigma^{vv.uM}$.
\end{pf}

It is instructive to check the relations of the parameter matrices in
\eqref{HuN} and~\eqref{Wuu} to regression coefficients and to
conditional covariance matrices.
With
$\Pi_{u|R}=-D_{uu}^{-1}(D_{uv},\break D_{uq}, D_{uC})$, one may write
\[
-D_{uu}\Pi_{u|v.C}=D_{uv}+D_{uq}\Pi_{q|v.C},\qquad
D_{uu}(Y_{u|C}-\Pi_{u|v.C}Y_{v|C})=D_{uu}Y_{u|vC},
\]
and for $W_{uu}$ defined in \eqref{recreg}
and specialized in \eqref{Wuu}
\[
D_{uu}^{-1}W_{uu}D_{uu}^{-T} =\Sigma_{uu|vqC}+\Pi_{u|q.vC}
\Sigma_{qq|vC}\Pi_{u|q.vC}^{\mathrm{T}}=\Sigma_{uu|vC},
\]
so that the required covariance matrix of $Y_{u|vC}$ is obtained.

The summary graph in node set $N$, induced by the generating parent
graph in node set~$V$, results now directly with Lemma~\ref{l5} applied to
equations \eqref{HuN} and \eqref{Wuu}, as is stated in Corollary~\ref{c4}.

\begin{coro}[(Generating the edge matrix of $\bolds{G}^{\bolds{V\setminus[C,M]}}_{\mathbf{sum}}$ from the edge
matrix
of a parent graph)]\label{c4} With the partially closed edge matrices corresponding to
Proposition~\ref{p4} and arranged in the appropriate order
\[
{\mathcal{D}}=\operatorname{zer}_{p}\tilde{{\mathcal{A}}},\qquad
\operatorname{zer}_q  \tilde{{\mathcal{S}}}^{FF.O}=
\pmatrix{{\mathcal{S}}_{qq|vC} & {\mathcal{P}}_{q|v.C} \cr
\cdot& {\mathcal{S}}^{vv.qO}
}
,
\]
the induced edge matrix components of the summary graph $G^{V\setminus
[C,M]}_{\mathrm{sum}}$ are
%
\begin{eqnarray} \label{cHuN} {\mathcal{H}}_{uu}&=&{\mathcal{D}}_{uu},
\qquad{\mathcal{H}}
_{uv}=\operatorname{In}[{\mathcal{D}}_{uv}+{\mathcal
{D}}_{uq}{\mathcal{P}}_{q|v.C} ], \qquad{\mathcal{S}}^{vv.uM},
\\ \label{cWuu}
{\mathcal{W}}_{uu}&=&\operatorname
{In}[({\mathcal{I}}_{uu}+{\mathcal{D}}
_{up}{\mathcal{D}}_{up}^{\mathrm{T}})+({\mathcal{D}}_{uq} {\mathcal
{S}}_{qq|vC} {\mathcal{D}}_{uq}^{\mathrm{T}})].
\end{eqnarray}
\end{coro}

\subsection{Non-Gaussian models associated with summary graphs}\label{sec3.3}

As noted before, the density $f_{N|C}$ of $Y_{N}$ given $Y_{C}$ is well
defined since it is obtained from a density of $Y_V$ generated over a
parent graph by marginalising over $Y_M$ and conditioning on $Y_C$. As
we have seen, this leads to the factorization of $f_{N|C}$ into $f_{u|v
C}$ and $f_{v |C}$. The independence structure of $Y_{v}$ given $Y_C$
is captured by
a concentration graph.

Corresponding models for discrete and continuous random variables have
been studied by Lauritzen and Wermuth (\citeyear{LauWer89}), extending the Gaussian covariance
selection models and the graphical, log-linear interaction models for
discrete variables. Maximum likelihood estimation is considerably
simplified for variation-independent parameters; see \citet{FrydLau89}.

For a joint Gaussian density $f_V$, the induced density $f_{u|v C}$ is
again Gaussian,
but in general, the form and parametrization of the density $f_{u|v C}$
induced by $f_V$ may be complex.
Nevertheless, we conjecture that the parameters associated with
$G^{V\setminus[C,M]}_{\mathrm{sum}}$ may often be obtained via
the notional
stepwise generating process described in Section~\ref{sec1.3}, that is, by
introducing latent variables that are mutually independent and
independent of $Y_v, Y_C$.

If the additional latent variables
are taken to be discrete and to have a large number of levels, then it
should be possible to generate, or at least to approximate closely
enough, any association corresponding to $i \dal k$ that does not
depend systematically on third variables. For discrete variables, this
follows with Theorem 1 of \citet{HollRos89} and otherwise presumably
by using Proposition 5.8 of \citet{Stu05}, but a proof is
pending.

\subsection{Generating a summary graph from a larger summary graph}\label{sec3.4}
Let a summary graph in node set $N'$ be given, where the corresponding
model, actually or only notionally, arises
from a parent graph model by conditioning on $Y_c$
and by marginalising over variables $Y_m$.

Then, the starting linear parent graph model is the triangular system
of equation \eqref{lsem2} in a mean-centered Gaussian variable $Y$ where
\[
AY=\varepsilon, \qquad\operatorname
{cov}(\varepsilon)=\Delta\mbox{ diagonal},
\qquad
A\mbox{ unit upper-triangular}.
\]
With Proposition~\ref{p4}, one obtains for $V\setminus\{c,m\}=(\mu
,\nu)$ the following equations in $Y_{\mu|c}$, $Y_{\nu|c}$, which
coincide in form with equation \eqref{linsumm} with $H^{\prime}_{\nu
N^{'}}=B_{\nu N^{'}}$
%
\begin{equation} \label{summc1}
\pmatrix{ B_{\mu\mu} & B_{\mu\nu} \cr
0 & \Sigma^{\nu\nu.\mu m}
}
\pmatrix{ Y_{\mu|c}\cr
Y_{\nu|c}
}
=
\pmatrix{ \eta'_\mu\cr
\zeta_{\nu}
}
, \qquad\operatorname{cov}
\pmatrix{ \eta'_\mu\cr
\zeta_{\nu}
}
=
\pmatrix{ W_{\mu\mu}'
& 0\cr
\cdot& \Sigma^{\nu\nu.\mu m}
}
.
\end{equation}

With added conditioning on a set $c_\nu\subseteq\nu$, no additional
ancestors of $c_\nu$ are defined, since every node in $\nu$ is
already an ancestor of $c$. But, with added conditioning on $c_\mu
\subseteq\mu$, the set $\mu\setminus c_\mu$ is split into foster
nodes $f_\mu$ of $c_\mu$ and into outsiders $o$ of $\{r, \nu\}$,
where $r=\{c_\mu, f_\mu\}$.

The equations for $Y_\mu$ are always block-triangular
in $(o,r)$. But, by contrast to the split of $V$ into $(O,R)$ in
equation~\eqref{OSmat}, these equations are not orthogonal in $(o,r)$
so that conditioning on $c_\mu$ in the summary graph is more complex
than conditioning directly on a set in the parent graph.

\begin{prop}[(Linear equations obtained from
\eqref{summc1} after conditioning on $\bolds{Y_{c_\mu}}$, $\bolds{Y_{c_\nu}}$ and
marginalising over $\bolds{Y_h}$, $\bolds{Y_l}$)]\label{p5}
Given \eqref{summc1} to $G^{V\setminus\{c,m\}}_{\mathrm{sum}}$,
where $o$ contains all outsiders of $\{c_\mu, f_\mu, \nu\}$,
equations for $Y_{\mu}$ are block-triangular in
\[
\mu=(o, r), \qquad\mbox{where } r=\{c_\mu, f_\mu\}.
\]
The additional
conditioning set $\{c_\mu,c_\nu\}$,
and additional marginalising sets $h\subseteq o$
and $l \subseteq\{f_\mu, \nu\setminus c_\nu\}$ give $C=\{c, c_\mu,
c_\nu\}$ and $M=\{m,h,l\}$. With $\psi=(r,\nu),$
the new equations are block-triangular and orthogonal in $(u,v)$, where
\[
u=o\setminus h, \qquad\phi=\psi\setminus\{
c_\mu, c_\nu\},\qquad
v=\phi\setminus l.
\]
With orthogonalised residuals $\xi_o=\eta'_{o}-Q_{or} \eta'_r$,
orders $\mu=(h,u,r)$, $\phi=( l, v)$ and
\[
Q_{\mu\mu}=\operatorname{inv}_r
\tilde{W'}_{\mu\mu},\qquad C_{o\psi}=B_{o\psi}-Q_{or}B_{r
\psi},
\qquad
K=\operatorname{inv}_{hl}
\pmatrix{ \tilde{B}_{oo} & \tilde{C}_{o\phi}
\cr
 0 & \tilde{\Sigma}^{\phi\phi.om}
},
\]
the linear summary graph model to $G^{V\setminus[C,M]}_{\operatorname
{sum}}$ is
%
\begin{equation} \label{sumuv}
\pmatrix{ K_{uu} & K_{uv}
\cr
 0 & \Sigma^{vv.uM}
}
\pmatrix{ Y_{u|C}\cr
Y_{v|C}
}
=
\pmatrix{ \eta_u\cr \zeta_{v}
}
, \qquad\eta_u=\xi_u - K_{uh}\xi_h - K_{ul}\Sigma_{ll|vC}
\zeta_l,
\end{equation}
and coincides with the linear model obtained from the triangular system
\eqref{lsem2} by directly conditioning on $Y_C$ and marginalising over
$Y_M$.
\end{prop}

\begin{pf}
The conditioning set $c_\mu$ splits the set of nodes $\mu$ into
$(o,r)$, where $o$ is without any descendant in $r=\{c_\mu, f_\mu\}$
and every node in $f_\mu$ has a descendant in $c$. This implies a~%
block-triangular form of $B_{\mu\mu}$ in $(o,r)$ in the equations of
$Y_{\mu|\nu c}$, however, with correlated residuals $\eta'_o$ and
$\eta'_r$.

For $\psi=(r,\nu)$, block-orthogonality with respect to $(o,\psi)$
in the
equations in $Y_{o|c}$ and $Y_{\psi|c}$
is achieved by subtracting from $\eta'_o$ the value predicted by
linear least-squares regression of $\eta'_o$ on
$\eta'_r $ and $\zeta_\nu$. This reduces, because of the
orthogonality of the equations in $(\mu,\nu)$, to subtracting $Q_{or}
\eta'_r$ from $\eta'_o$.

The matrix of equation parameters of $Y_{\psi|c}$ coincides with
the concentration matrix of $Y_{\psi|c}$
given by
%
\begin{equation}\label{newcon}\Sigma^{\psi\psi.om} = \Sigma_{\psi
\psi|c}^{-1} =
\pmatrix{ B_{rr}^{\mathrm{T}} Q_{rr}B_{rr}&
B_{rr}^{\mathrm{T}} Q_{rr} B_{r\nu}\cr
\cdot & \Sigma^{-1}_{\nu\nu|c}+B_{r\nu}^{\mathrm{T}} Q_{rr} B_{r\nu}
}
.
\end{equation}

By the block-triangularity and orthogonality in $(o, \psi)$, the
equations in $Y_{o|c}$ can be replaced by equations
in $Y_{o|C}$.
For the equations in $Y_{\phi|C}$, the matrix of equation parameters is
$\Sigma_{\phi\phi|C}^{-1}=[\Sigma_{\psi\psi|c}^{-1}]_{\phi,\phi
}=\Sigma^{\phi\phi.o m}$.
The resulting equations give the Gaussian linear model
to the summary graph in node set $V\setminus\{C, m\}=(o,\phi)$.

In the linear model to $G^{V\setminus[C, m]}$, marginalising over
$Y_{h|C}$, where $h \subseteq o$, and over $Y_{l|C}$, where $l
\subseteq\phi$, is achieved by partial inversion on $h,l$ of the
block-triangular matrix of equation parameters
and by keeping only the equations in $Y_{u|C}$ and $Y_{v|C}$.

In the resulting equation \eqref{sumuv}, one knows by the
commutativity and exchangeability of partial inversion for $m=(g,k)$,
$p=\{g,h\}$, $q=\{k,l\}$ that
\[
K_{uu}=[\operatorname{inv}_h \operatorname{inv}_g A]_{u,u}=
[\operatorname{inv}_p A]_{u,u},
\]
so that $K_{uu}=D_{uu}$, where $D$ is defined for Proposition \ref{p4}.
Furthermore, by the properties of reduced form equations
\[
- K_{uu}\Pi_{u|v.C}= K_{uv}=D_{uv}+D_{uq}\Pi_{q|v.C},
\]
so that the parameter matrices of $Y_{u|C}$ and $Y_{v|C}$ given in
\eqref{sumuv} coincide with those
in \eqref{HuN} and \eqref{Wuu} of Proposition \ref{p4} -- that
is, they give the Gaussian linear model
to the summary graph in node set $V\setminus\{C, N\}=(u,v)$.
\end{pf}

Since partial closure has the same exchangeability property as partial
inversion and both operators are commutative, the same type of proof
holds for the edge matrix expression corresponding to \eqref{sumuv}.

\begin{coro}[(Generating the edge matrix of $\bolds{G}^{\bolds{V\setminus[C,M]}}_{\mathbf{sum}}$ from the edge
matrix
of a summary graph)]\label{c5}
For $c\subset C$ and $m \subset M$, edge matrix components of the
summary graph $G^{V\setminus[C,M]}$ result from the edge matrix
components $\mathcal{B}_{\mu\mu}$, $\mathcal{B}_{\mu\nu}$,
${\mathcal{W}}'_{\mu mu}$
and ${\mathcal{S}}^{\nu\nu.\mu m}$ of $G^{V\setminus[c,m]}$ by
using the
transformed edge matrices
\[
{\mathcal{Q}}_{\mu\mu}=\operatorname{zer}_r \tilde{{\mathcal
{W}}'}_{\mu\mu}, \qquad{\mathcal{C}}
_{o\psi} =\operatorname{In}[\mathcal{B}_{o\psi}+ {\mathcal
{Q}}_{or}\mathcal{B}_{r\psi}],\qquad{\mathcal{K}}
=\operatorname{zer}_{hl}
\pmatrix{ \tilde{\mathcal{B}}_{oo} & \tilde{{\mathcal
{C}}}_{o\phi}
\cr 0 & \tilde{{\mathcal{S}}}^{\phi\phi.om}
}
\]
to obtain ${\mathcal{K}}_{uu}$, ${\mathcal{K}}_{uv}$ directly,
${\mathcal{S}}^{vv.uM}$ as
the edge matrix to \eqref{newcon}, and
%
\begin{equation}\label{calwuu}
{\mathcal{W}}_{uu}=\operatorname{In}[{\mathcal
{Q}}_{uu}+ {\mathcal{K}}_{u h}{\mathcal{Q}}
_{hh}{\mathcal{K}}_{u h}^{\mathrm{T}}+{\mathcal{K}}_{ul} {\mathcal{S}}_{ll|v
C}{\mathcal{K}}_{ul}^{\mathrm{T}}].
\end{equation}
\end{coro}

\subsection{Path results derived from edge matrix
transformations}\label{sec3.5}

If one starts with the summary graph $G^{V\setminus
[c,m]}_{\mathrm{sum}}$ and
conditions by using Corollary~\ref{c5}, edges are induced by $r$-line
collision paths, where we let
$r=\{c_\mu,f_\mu\}=\{\raisebox{-1pt}{\mbox{\includegraphics{309i01.eps}}}\}$:
\begin{longlist}[(a)]
\item[(a)]
$\mbox{\Large{$\mbox{$\circ$}$}}_\mu
\dal
\mbox{\Large{$\mbox{$\circ$}$}}_\mu$
\mbox{ results with }
$\mbox{\Large{$\mbox{$\circ$}$}}_\mu
\dal
\raisebox{-1pt}{\mbox{\includegraphics{309i01.eps}}}
\cdots
\raisebox{-1pt}{\mbox{\includegraphics{309i01}}}
\dal
\mbox{\Large{$\mbox{$\circ$}$}}_\mu$,

\item[(b)] \hspace*{-0,69pt}
$\mbox{\Large{$\mbox{$\circ$}$}}_\psi
\ful
\mbox{\Large{$\mbox{$\circ$}$}}_\psi$
\mbox{ results with }
$\mbox{\Large{$\mbox{$\circ$}$}}_\psi
\fra
\raisebox{-1pt}{\mbox{\includegraphics{309i01.eps}}}
\dal
\raisebox{-1pt}{\mbox{\includegraphics{309i01}}}
\cdots
\raisebox{-1pt}{\mbox{\includegraphics{309i01}}}
\dal
\raisebox{-1pt}{\mbox{\includegraphics{309i01}}}
\fla
\mbox{\Large{$\mbox{$\circ$}$}}_\psi$,

\item[(c)]
$\mbox{\Large{$\mbox{$\circ$}$}}_\mu
\fla
\mbox{\Large{$\mbox{$\circ$}$}}_\psi$
\mbox{ results with }
$\mbox{\Large{$\mbox{$\circ$}$}}_\mu
\dal
\raisebox{-1pt}{\mbox{\includegraphics{309i01.eps}}}
\cdots
\raisebox{-1pt}{\mbox{\includegraphics{309i01}}}
\dal
\raisebox{-1pt}{\mbox{\includegraphics{309i01}}}
\fla
\mbox{\Large{$\mbox{$\circ$}$}}_\psi$.
\end{longlist}
The corresponding relevant edge matrix expressions are, respectively,
$Q_{\mu\mu}=\operatorname{zer}_r W_{\mu\mu}$, $\operatorname{In}[
B_{r \psi}^{\mathrm{T}} Q_{rr} B_{r
\psi}]$ and
$\operatorname{In}[ {\mathcal{Q}}_{or}\mathcal{B}_{r\psi}] $. For
each pair, one keeps one edge
of several of the same kind. The subgraph induced by nodes $(o, \phi)$
is $G^{V\setminus[C, m]}$.

By marginalising next over $m'=(h,l)=$(\raisebox{-2pt}{$\mbox{\includegraphics{309i02.eps}}_{h}$},
\raisebox{-2pt}{$\mbox{\includegraphics{309i02}}_{l}$}) in
the graph $G^{V\setminus[C,m]}_{\mathrm{sum}}$, three
types of edges are
induced when closing $m'$-line transmitting paths:
\begin{longlist}[(a)]
\item[(d)] \hspace*{-0,69pt}
$\mbox{\Large{$\mbox{$\circ$}$}}_\phi
\ful
\mbox{\Large{$\mbox{$\circ$}$}}_\phi$
\mbox{ results with }
$\mbox{\Large{$\mbox{$\circ$}$}}_\phi
\ful$
\raisebox{-2pt}{$\mbox{\includegraphics{309i02.eps}}_l$}
$\cdots$
\raisebox{-2pt}{$\mbox{\includegraphics{309i02}}_l$}
$\ful
\mbox{\Large{$\mbox{$\circ$}$}}_\phi$,

\item[(e)]
$\mbox{\Large{$\mbox{$\circ$}$}}_o
\fla
\mbox{\Large{$\mbox{$\circ$}$}}_o$
\mbox{ results with }
$\mbox{\Large{$\mbox{$\circ$}$}}_o
\fla$
\raisebox{-2pt}{$\mbox{\includegraphics{309i02.eps}}_h$}
$\cdots$
\raisebox{-2pt}{$\mbox{\includegraphics{309i02}}_h$}
$\fla
\mbox{\Large{$\mbox{$\circ$}$}}_o$,

\item[(f)]
$\mbox{\Large{$\mbox{$\circ$}$}}_o
\fla
\mbox{\Large{$\mbox{$\circ$}$}}_\phi$
\mbox{ results with }
$\mbox{\Large{$\mbox{$\circ$}$}}_o
\fla$
\raisebox{-2pt}{$\mbox{\includegraphics{309i02.eps}}_h$}
$\cdots$
\raisebox{-2pt}{$\mbox{\includegraphics{309i02}}_h$}
$\fla$
\raisebox{-2pt}{$\mbox{\includegraphics{309i02}}_l$}
$\ful$
\raisebox{-2pt}{$\mbox{\includegraphics{309i02}}_l$}
$\cdots$
\raisebox{-2pt}{$\mbox{\includegraphics{309i02.eps}}_l$}
$\ful
\mbox{\Large{$\mbox{$\circ$}$}}_\phi$,

\item[(g)]   \hspace*{-0,69pt}
$\mbox{\Large{$\mbox{$\circ$}$}}_u
\dal
\mbox{\Large{$\mbox{$\circ$}$}}_u$
\mbox{ results with }
$\mbox{\Large{$\mbox{$\circ$}$}}_u
\fla$
\raisebox{-2pt}{$\mbox{\includegraphics{309i02.eps}}_h$}
$\dal$
\raisebox{-2pt}{$\mbox{\includegraphics{309i02}}_h$}
$\fra
\mbox{\Large{$\mbox{$\circ$}$}}_u$,

\item[(h)]   \hspace*{-0,69pt}
$\mbox{\Large{$\mbox{$\circ$}$}}_u
\dal
\mbox{\Large{$\mbox{$\circ$}$}}_u$
\mbox{ results with }
$\mbox{\Large{$\mbox{$\circ$}$}}_u
\fla$
\raisebox{-2pt}{$\mbox{\includegraphics{309i02.eps}}_l$}
$\ful$
\raisebox{-2pt}{$\mbox{\includegraphics{309i02}}_l$}
$\fra
\mbox{\Large{$\mbox{$\circ$}$}}_u$.
\end{longlist}
The corresponding edge matrix expressions are, respectively,
$ {\mathcal{K}}_{\phi\phi}$, ${\mathcal{K}}_{oo}$, $ {\mathcal
{K}}_{o\phi}$,\break $\operatorname{In}[{\mathcal{K}}_{uh}{\mathcal{Q}}_{hh}{\mathcal{K}}_{u h}^{\mathrm{T}}]$ and $\operatorname
{In}[{\mathcal{K}}_{ul} {\mathcal{S}}_{ll|v
C}{\mathcal{K}}_{ul}^{\mathrm{T}}]$.
After keeping just one edge of several of the same kind, the subgraph
induced by nodes $(u, v)$ is $G^{V\setminus[C, M]}$.

Notice that the effect of the indicator function is to reduce several
edges of the same kind to just one. The closed form expressions
of the edge matrix results imply that some of the paths are to be
closed in the given order.

The edge matrices $\operatorname{In}[ {\mathcal{Q}}_{or}\mathcal
{B}_{r\psi}] $ and ${\mathcal{K}}
_{o\phi}$ correspond in a Gaussian
summary graph model to orthogonalising, that is, to removing some
residual correlations.
By the associated steps, (c) or (f), $ik$-arrows may be generated for
which node $k$ is not an ancestor of $i$ in the generating graph.

In contrast, for the outsiders of the conditioning set, such as set $o$
in the summary graph in nodes $(o, \phi)$, there is an $ik$-arrow
if and only if $k$ is a parent or a forefather of node $i$ in the
larger generating parent graph because the only arrow-inducing paths
for the subset $o$ are those in (e).

Since a summary graph results after conditioning with steps (a)--(c)
and also after marginalising with steps (d)--(h), summary graphs are
said to be closed under marginalising and conditioning and one may
reverse the order of conditioning and marginalising. The following
example illustrates such reversed stepwise constructions.

\begin{figure}[b]

\includegraphics{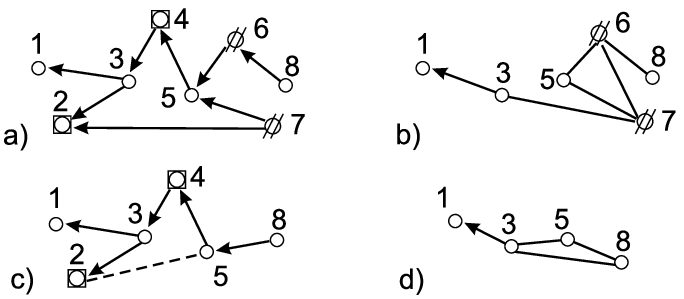}

\caption{\textup{(a)} The generating graph $G^{V}_{\operatorname{par}}$, \textup{(b)}
$G^{V\setminus[C, \varnothing]}_{\mathrm{sum}}$, \textup{(c)} $G^{V\setminus
[\varnothing,
M]}_{\mathrm{sum}}$, \textup{(d)} $G^{N}_{\mathrm{sum}}$.}
\label{fig7}
\end{figure}

\begin{exa}[(Path constructions of $\bolds{G}^{\bolds{V\setminus
[C,M]}}_{\mathbf{sum}}$
for $\bolds{M=q}$ and $\bolds{p=}\bm{\varnothing}$)]\label{e5}
The node
set of the parent graph is $V=(1, \dots, 8)$. The conditioning set is
$C=\{2,4\}$ and the marginalising set is $M=\{6,7\}$. The foster nodes
of $C$, are in $F=\{3,5,6,7,8\}$ and $u=O=\{1\}$,
$v=\{3,5,8\}$.
\end{exa}

In this example with graphs in Figure~\ref{fig7}, the summary graph model
is equivalent to a~triangular system in $N=(1,3,5,8)$ even though
$G^{V\setminus[\varnothing, M]}_{\mathrm{sum}}$ is not Markov
equivalent to any
directed acyclic graph since it contains the chordless collision path
$3\fra2\dal5\fla8$.
It is typical that further marginalising or conditioning may again lead
to simpler graphs and models.

With just one node in the marginalising set, the paths (d)--(h) have
just two edges. In addition, by the properties of partial inversion
and partial closure, the paths (a)--(c) can be closed by
repeatedly closing paths of just two edges.
This leads to operating on one node at a time in any order; see
also the \hyperref[app]{Appendix}, Table~\ref{tab1} and Proposition~\ref{p1}.

\subsection{The MAG corresponding to $G^{V\setminus
[C,M]}_{\mathrm{sum}}$ and
local Markov properties}\label{sec3.6}

The keys to deriving the MAG corresponding to $G^{V\setminus
[C,M]}_{\mathrm{sum}}
$ are the definition of the variables in the Gaussian MAG model and the
result \eqref{zerprod}.
For $Y_v$, the summary graph and the MAG specify the same concentration
graph, and dependences to arrows pointing from $v$ to $u$ also coincide.

A full order of the nodes in $u$ of $G^{V\setminus[C,M]}_{\operatorname
{sum}}$ may
sometimes be given by the arrows, such as in Figure~\ref{fig3}(b).
Sometimes there is none, as in Figure~\ref{fig2}(b). More often there is a
partial order, such as in Figure~\ref{fig1}(d) or~\ref{fig7}(c). Then one may
take any compatible full ordering of the nodes in $u$ in which the
ancestors within $u$ of each node $i$ in $G^{V\setminus
[C,M]}_{\mathrm{sum}}$
are in the past of $i$, that is, in $\{i+1, \ldots, d_u\}$.

For each node $i$, we let $c_i\subseteq\{i+1, \ldots, d_u\}$ denote
the ancestors of $i$ in $G^{V\setminus[C,M]}_{\mathrm{sum}}$ and
$\bar{c}_{i}=
\{i+1, \ldots, d_u\}\setminus c_i$. Next, we derive for each node pair
$i,k$ with $k$ in $c_i$ and each node pair $i,l$ with $l$ in $\bar{c}_i$,
the edges in the MAG corresponding to $G^{V\setminus
[C,M]}_{\mathrm{sum}}$ by
applying \eqref{zerprod} to equation~\eqref{recreg}.

For $a=( 1, \ldots, i, \bar{c}_i )$ and $b=c_i$, the vector
${\mathcal{P}}_{i|b}=\operatorname{In}[{\mathcal{K}}_{ib} +{\mathcal
{Q}}_{ib}{\mathcal{K}}_{bb}]$ gives zeros
and ones for the dependence of $Y_i$ on $Y_{c_i}$
given $Y_v$, $Y_C$ and\vspace*{2pt}
%
\begin{equation}\label{rep1}
 \mbox{in the MAG, } i\fla k \mbox{ for }
\operatorname{In}[{\mathcal{P}}
_{i|k.b\setminus k}]=1,\qquad
i,k \mbox{ uncoupled, otherwise}.\vspace*{2pt}
\end{equation}

Similarly, for $i,l$ we let $e_{il}=c_i \cup c_l$ and $\bar{e}_{il}=\{
i+1, \ldots, d_u\}\setminus e_{il}$, take $ a=( 1, \ldots, i , l,
\bar{e}_{il})$
and $b=e_{il}$. With ${\mathcal{S}}_{aa|b}=\operatorname
{In}[{\mathcal{K}}_{aa}{\mathcal{Q}}_{aa}{\mathcal{K}}
_{aa}] $ of \eqref{zerprod}, ${\mathcal{K}}_{il}=0$ and ${\mathcal
{W}}_{uu}$ the
edge matrix\vadjust{\goodbreak} of the covariance graph of $G^{V\setminus
[C,M]}_{\mathrm{sum}}$:
%
\begin{equation}\label{rep2}
 \mbox{in the MAG, } i\dal l \mbox{ for }
\operatorname{In}[
{\mathcal{W}}_{il.b}]=1,\qquad
i,l \mbox{ uncoupled, otherwise}.
\end{equation}
The corresponding MAG results after inserting or replacing edges in
$G^{V\setminus[C,M]}_{\mathrm{sum}}$ according to \eqref{rep1}
and \eqref
{rep2} and keeping just one of several same edges.

\begin{prop}[(Local Markov properties of summary
graphs)]\label{p6} Let the edge matrix components, $H_{uN}$, $W_{uu}$ and
${\mathcal{S}}^{vv.uM}$ of $G^{V\setminus\{C,M\}}_{\mathrm{sum}}$
be given from Corollary~\ref{c5}. Let node $l$ and sets $c_i, e_{il}$ be
defined as above, but their subscripts dropped. Let further $\beta$
denote subsets of nodes uncoupled to node $i$, then:
\begin{enumerate}[(1)]
\item[(1)]$i \perp\!\!\!\hspace*{-1pt}\perp\beta|C v\setminus\{i,\beta\} \iff
{\mathcal{S}}^{i\beta
.uM}=0$ for $i\in v$ and $\beta\subset v$.

\item[(2)] $i\perp\!\!\!\hspace*{-1pt}\perp\beta|C v\setminus\beta\iff{\mathcal
{H}}_{i\beta.c}=0$
for $i\in u$ and $\beta\subset v$.

\item[(3)] $ i\perp\!\!\!\hspace*{-1pt}\perp l|Cve \iff({\mathcal{W}}_{il}=0$ and $
{\mathcal{W}}_{ie}^{\hspace*{.35em}
}{\mathcal{W}}_{ee}^{-}{\mathcal{W}}_{el}=0)$ for $i \in u$, and
$l\in\bar{c}$.

\item[(4)] $i\perp\!\!\!\hspace*{-1pt}\perp\beta|Cvc\setminus\beta\!\iff\!
({\mathcal{H}}_{i \beta
}=0$ and ${\mathcal{W}}_{ic}{\mathcal
{W}}_{cc}^{-}{\mathcal{H}}_{c\beta}=0)$
for $i \in u\!$ and $ \beta\subset c$.
\end{enumerate}
\end{prop}

Notice that pairwise independences result if $\beta$'s contain single
elements.

\begin{pf*}{Proof of Proposition~\ref{p6}}
The independences in (1) within $v$ are those of a concentration
graph; see also \eqref{indMRC} in Example~\ref{e4}. The independences in (2)
are those
obtained when regressing $Y_{i|C}$ on $Y_{v|C}$; see also Example~\ref{e2}.
The independences in (3) and (4) are reformulations of \eqref{rep2}
and \eqref{rep1}, respectively.
\end{pf*}

\section{Discussion}\label{sec4}

The common attractive feature of a maximal ancestral graph and of the
corresponding summary graph is that they elucidate consequences of a
possibly much larger generating graph regarding independences. The
smaller graphs capture the independence structure implied by the
generating graph and
they can be used to understand additional consequences of the
generating graph for independences that result after additional
marginalising and conditioning.

An advantage of the MAG is that each edge corresponds to a conditional
association, each missing edge to a conditional independence. A
disadvantage of a MAG is that a dependence, say to $i\fla k$, may be
severely distorted
compared to the dependence to $i\fla k$ in the generating process.
With the corresponding summary graph, one can identify which of the
conditional dependences in the MAG remain undistorted and which do not.

Given the summary graph, the corresponding MAG is derived in a few
steps. But in general, one cannot obtain from a given MAG the
corresponding summary graph or the information about distortions. Both
types of graph may contain semi-directed cycles. These are typically of
interest only in connection with a larger generating process.

In contrast, their common subclass of
regression graphs gives a substantial and much needed enlargement of
the types of research hypotheses that can be formulated with directed
acyclic graphs. They model stepwise generating processes not only in
univariate but also in joint responses. This leads to a corresponding
recursive factorization of the joint density in these vector variables.

In addition, every independence constraint
for a component of a joint response is conditional on variables in the
past of the joint response.
This is an important distinction from all other types of currently
known chain graphs and is in line with research in many substantive fields
where the study of dependences on past variables is judged to be more
fruitful than those of associations and of independences among
variables arising at the same time.

For Gaussian regression graph models, properties of estimators and test
statistics have been quite well understood for a considerable time.
For discrete random variables, all regression graph models are smooth;
see \citet{Drton09}. Such smooth models are curved exponential
families (see \citet{Cox06}, Section 6.8) so that they have desirable
properties regarding estimation and asymptotic properties of tests.

Much less is known
for joint responses of
discrete and continuous random components. Thus, though we now can
derive important consequences of any type of regression graph model,
more results on equivalence, identification, estimation and
goodness-of-fit criteria are needed.

However, if the regression graph model can be generated, as discussed,
via special types of hidden variables in a larger parent graph model,
then its independence
structure is defined by a list of independence statements for variable
pairs. This permits local fitting with univariate generalized linear
models, with checks for
linearity, interaction and conditional independence based on observed
associations of variable pairs and triples.

This requires no knowledge about the form of the joint distribution and
it permits us to formulate research hypotheses that are compatible with
a given set of data and that are to be investigated in further
empirical studies.

\begin{appendix}
\section*{Appendix: Two-edge paths of summary graphs}\label{app}

The following
arguments show that the types of induced edges of Table~\ref{tab1} are self-consistent:
A~node to be marginalised over is again denoted by
$\raisebox{-2pt}{\mbox{\includegraphics{309i02.eps}}}$
and a node to be conditioned on by $\raisebox{-1pt}{\mbox{\includegraphics{309i01.eps}}}$.

\renewcommand{\thefigure}{\arabic{figure}}
\setcounter{figure}{7}
\begin{figure}[t]

\includegraphics{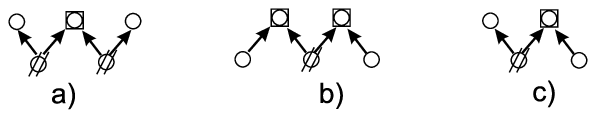}

\caption{Active alternating paths that generate two-edge paths
\textup{(a)} of type (4) inducing
$\mbox{\Large{$\mbox{$\circ$}$}}\dal\mbox{\Large{$\mbox{$\circ$}$}}$, \textup{(b)} of type (5) inducing
$\mbox{\Large{$\mbox{$\circ$}$}}\ful\mbox{\Large{$\mbox{$\circ$}$}}$,
\textup{(c)} of type (6) or (7) inducing $\mbox{\Large{$\mbox{$\circ$}$}}\fla\mbox{\Large{$\mbox{$\circ$}$}}$.}
\label{fig8}
\end{figure}

The three types of edge-inducing, two-edge paths (1)--(3) in a parent
graph that have as an inner node a transition, a source or a sink node,
respectively, are
defined to generate the following three different types of edges:
\begin{enumerate}[(1)]
\item[(1)]
$\mbox{\Large{$\mbox{$\circ$}$}}
\fla
\raisebox{-2pt}{\mbox{\includegraphics{309i02.eps}}}
\fla
\mbox{\Large{$\mbox{$\circ$}$}}
\Longrightarrow
\mbox{\Large{$\mbox{$\circ$}$}}
\fla
\mbox{\Large{$\mbox{$\circ$}$}}$,

\item[(2)]
$\mbox{\Large{$\mbox{$\circ$}$}}
\fla
\raisebox{-2pt}{\mbox{\includegraphics{309i02.eps}}}
\fra
\mbox{\Large{$\mbox{$\circ$}$}}
\Longrightarrow
\mbox{\Large{$\mbox{$\circ$}$}}
\dal
\mbox{\Large{$\mbox{$\circ$}$}}$,

\item[(3)]
$\mbox{\Large{$\mbox{$\circ$}$}}
\fra
\raisebox{-1pt}{\mbox{\includegraphics{309i01.eps}}}
\fla
\mbox{\Large{$\mbox{$\circ$}$}}
\Longrightarrow
\mbox{\Large{$\mbox{$\circ$}$}}
\ful
\mbox{\Large{$\mbox{$\circ$}$}}$.
\end{enumerate}
The arrow has one, the dashed line two and the full line no edge
endpoints that define a collision node
when the edge is mirrored at the same node.
Dashed lines denote edges in covariance graphs and full lines in
concentration graphs.
Closing paths in such graphs are defined to preserve the type of edge:
\begin{enumerate}[(4)]
\item[(4)]
$\mbox{\Large{$\mbox{$\circ$}$}}
\dal
\raisebox{-1pt}{\mbox{\includegraphics{309i01.eps}}}
\dal
\mbox{\Large{$\mbox{$\circ$}$}}
\Longrightarrow
\mbox{\Large{$\mbox{$\circ$}$}}
\dal
\mbox{\Large{$\mbox{$\circ$}$}}$,

\item[(5)]
$\mbox{\Large{$\mbox{$\circ$}$}}
\ful
\raisebox{-2pt}{\mbox{\includegraphics{309i02.eps}}}
\ful
\mbox{\Large{$\mbox{$\circ$}$}}
\Longrightarrow
\mbox{\Large{$\mbox{$\circ$}$}}
\ful
\mbox{\Large{$\mbox{$\circ$}$}}$.
\end{enumerate}
The next two paths, (6) and (7), and both induce an
arrow:
\begin{enumerate}[(6)]
\item[(6)]
$\mbox{\Large{$\mbox{$\circ$}$}}
\dal
\raisebox{-1pt}{\mbox{\includegraphics{309i01.eps}}}
\fla
\mbox{\Large{$\mbox{$\circ$}$}}
\Longrightarrow
\mbox{\Large{$\mbox{$\circ$}$}}
\fla
\mbox{\Large{$\mbox{$\circ$}$}}$,

\item[(7)]
$\mbox{\Large{$\mbox{$\circ$}$}}
\fla
\raisebox{-2pt}{\mbox{\includegraphics{309i02.eps}}}
\ful
\mbox{\Large{$\mbox{$\circ$}$}}
\Longrightarrow
\mbox{\Large{$\mbox{$\circ$}$}}
\fla
\mbox{\Large{$\mbox{$\circ$}$}}$.
\end{enumerate}
Paths (4)--(7) arise from active alternating paths in a parent graph
for which inner source nodes in $\{\raisebox{-2pt}{\mbox{\includegraphics{309i02.eps}}}\}$
alternate with inner sink nodes in $\{\raisebox{-1pt}{\mbox{\includegraphics{309i01.eps}}}\}$:

The two-edge paths (4)--(7) result from Figure~\ref{fig8} as follows: path (4)
from (a) by only marginalising, path (5) from (b) by only conditioning,
path (6) from (c) by only marginalising and path (7) from (c) by only
conditioning. The paths (a)--(c) of Figure~\ref{fig8} generalize paths (2),
(3) and (1), respectively.

\begin{figure}[t]

\includegraphics{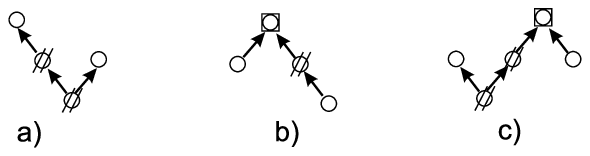}

\caption{Active paths that generate two-edge paths (a) of type
(8) inducing
$\mbox{\Large{$\mbox{$\circ$}$}}\dal\mbox{\Large{$\mbox{$\circ$}$}}$, (b) of type (9) inducing
 $\mbox{\Large{$\mbox{$\circ$}$}}\ful\mbox{\Large{$\mbox{$\circ$}$}}$,
and (c) of type (10) inducing $\mbox{\Large{$\mbox{$\circ$}$}}\fla\mbox{\Large{$\mbox{$\circ$}$}}$.}
\label{fig9}
\end{figure}

The three remaining edge-inducing paths of two edges in $G^{V\setminus
[C,M]}_{\mathrm{sum}}$ are
\begin{enumerate}[(10)]
\item[(8)]
$\mbox{\Large{$\mbox{$\circ$}$}}
\fla
\raisebox{-2pt}{\mbox{\includegraphics{309i02.eps}}}
\dal
\mbox{\Large{$\mbox{$\circ$}$}}
\Longrightarrow
\mbox{\Large{$\mbox{$\circ$}$}}
\dal
\mbox{\Large{$\mbox{$\circ$}$}}$,

\item[(9)]
$\mbox{\Large{$\mbox{$\circ$}$}}
\ful
\raisebox{-2pt}{\mbox{\includegraphics{309i02.eps}}}
\fla
\mbox{\Large{$\mbox{$\circ$}$}}
\Longrightarrow
\mbox{\Large{$\mbox{$\circ$}$}}
\ful
\mbox{\Large{$\mbox{$\circ$}$}}$,

\item[(10)]
$\mbox{\Large{$\mbox{$\circ$}$}}
\dal
\raisebox{-2pt}{\mbox{\includegraphics{309i02.eps}}}
\ful
\mbox{\Large{$\mbox{$\circ$}$}}
\Longrightarrow
\mbox{\Large{$\mbox{$\circ$}$}}
\fla
\mbox{\Large{$\mbox{$\circ$}$}}$.
\end{enumerate}
The three active paths of Figure~\ref{fig9} result by substituting the
undirected edges in (8)--(10) by the appropriate generating components
(2) or (3).

By marginalising over the transition node in Figure~\ref{fig9}(a)--(c), one generates, respectively,
path~(2), path (3) and the path in Figure~\ref{fig8}(c).

The construction of the summary graph simplifies considerably for
special types
of parent graphs -- for instance, for the graphs to the lattice
conditional independence
models studied by Andersson \textit{et al.} (\citeyear{Andetal97}), and for the graphs corresponding
to labeled trees, studied by Castelo  and Siebes (\citeyear{CastSieb03}).

\end{appendix}

\subsection*{Acknowledgements} I thank D.R. Cox, M. Drton, S.C. Lorin,
G.M. Marchetti and K. Sadeghi for their insightful and constructive
suggestions; the referees for searching questions; and P. Albin and M.
Studen\'{y} for sharing with me some of their knowledge on families of
complete distributions.
This paper is supported in part by the Swedish Research Society via the
Gothenburg Stochastic Center and by the Swedish Strategic Fund via the
Gothenburg Math. Modelling Center.

\printhistory

\end{document}